 \documentclass[aps,prd,superscriptaddress,showpacs,twocolumn,preprintnumbers]{revtex4}

\usepackage{epsfig,dcolumn}
\usepackage{graphicx}
\usepackage{color}

\usepackage{amsmath}
\usepackage{amsfonts}
\usepackage{amssymb}
\usepackage{graphicx}
\usepackage{type1cm}
\usepackage{eso-pic}
\newcommand{\be}{\begin{equation}}
\newcommand{\ee}{\end{equation}}

\newcommand{\im}{\mathrm{Im}\,}
\newcommand{\re}{\mathrm{Re}\,}

\newcommand{\degree}{{\rm o}}

\usepackage{bm} % scaled bold math symbols

\begin{document}

\title{The pion-kaon scattering amplitude constrained with forward dispersion relations up to 1.6 GeV} 

\author{J.R.~Pelaez}
\affiliation{Departamento de F\'isica Te\'orica II, Universidad Complutense de Madrid, 28040 Madrid, Spain}
\author{A.Rodas}
\affiliation{Departamento de F\'isica Te\'orica II, Universidad Complutense de Madrid, 28040 Madrid, Spain}
%\date{\today}

\begin{abstract}
In this work we provide simple and precise parameterizations of the existing $\pi K$ scattering data from threshold up
to 1.6 GeV, which are constrained to satisfy forward dispersion relations as well as three additional threshold sum rules. We also provide phenomenological values of the threshold parameters and of the resonance poles that appear in elastic scattering.
\end{abstract}
\maketitle

\section{Introduction}

Pion-kaon scattering is a very relevant process 
for our understanding of Hadron Physics and the strong interaction.
The motivation to study it is threefold.

First of all, because pions and/or kaons appear in the final states of all hadronic processes.
In particular kaons do so if the process involves net strangeness. Since pions and kaons 
interact strongly, final state $\pi K$ re-scattering effects are essential to describe and understand such hadronic processes. 

Second, the reaction is interesting by itself, because even though we cannot solve QCD at low energies, the identification of pions and kaons as pseudo-Goldstone Bosons of the QCD spontaneous chiral symmetry breaking allows 
for a rigorous formulation in terms of a low energy  effective theory known as 
 Chiral Perturbation Theory \cite{Gasser:1984gg} (ChPT). In turn, ChPT
provides $\pi K$ scattering amplitudes 
which have been calculated first to one-loop \cite{Bernard:1990kw} and then to two-loops \cite{Bijnens:2004bu}.
Relevant constraints on the ChPT low energy constants can be obtained from
sum rules and dispersion relations applied to $\pi K$ scattering \cite{Ananthanarayan:2001uy}.
In addition, $\pi K$ scattering was
subsequently unitarized to one-loop \cite{Dobado:1992ha,Pelaez:2004xp}
or within the chiral unitary approach \cite{Oller:1997ti}, providing a simultaneous description of the low-energy
and resonant regimes.
Moreover, there is a renewed interest in $\pi K$ scattering 
from Lattice QCD, where the 
main features, like threshold parameters \cite{Beane:2006gj},
scattering phases and resonances \cite{Lang:2012sv}, have already been calculated. Although the 
pion mass used for these lattice calculations is not physical,
one can expect physical values to be within reach soon.
Alternative lattice strategies that calculate 
$\pi K$ scattering from unitarized chiral 
Lagrangians have also been  followed recently in \cite{Doring:2011nd}.

Third, in pion-kaon scattering appear some of the still controversial light scalar mesons, like the $K_0^*(800)$ or $\kappa$ resonance and
the $K_0^*(1430)$. The former has been the subject of a longstanding debate about
its very existence and nature. 
Actually, it is a firm candidate to form the lightest nonet of scalar mesons together with the $f_0(500)$ or $\sigma-$meson, 
the $f_0(980)$ and the $a_0(980)$.  There are strong evidences that these states
might form
a nonet of non ordinary mesons \cite{Jaffe:1976ig,Oller:1997ti}, i.e., mesons not predominantly
made of a quark and an antiquark.
The $\kappa$ resonance has been obtained within 
different variants of
unitarized ChPT in 
\cite{Oller:1997ti,Pelaez:2004xp},
it has also been shown to have a mass smaller than 900 MeV \cite{Cherry:2000ut} 
and has been found \cite{DescotesGenon:2006uk} 
from a rigorous {\it solution} \cite{Buettiker:2003pp} of Roy-Steiner dispersion relations \cite{Steiner:1971ms}, which is the best determination so far.
However,  those pieces of evidence are still not considered enough
by the Review of Particle Properties (RPP) \cite{RPP}, which still lists the $K_0^*(800)$ resonance 
under the ``needs confirmation'' label.  Thus, the $\kappa$-meson is a further motivation
for our present study, since any rigorous resonance
determination from data 
(not a solution of dispersion relations or lattice) 
requires first a consistent knowledge of $\pi K$ scattering,
which, in order to control all uncertainties should reach beyond the pure elastic regime. Incidentally, the latter region is also of direct interest for the $K_0^*(1430)$ resonance.

Hence, the goal of this work is to perform an analysis of the existing $\pi K$ scattering data
constrained to satisfy Forward Dispersion Relations. 
The advantage of these relations is that,
contrary to other kinds of dispersion relations (like Roy-Steiner equations in their simplest form),
they can be easily implemented up to arbitrarily high energies. 
Here we will apply to $\pi K$ scattering
an approach that has been recently followed \cite{Pelaez:2004vs}
 to obtain a precise description of $\pi\pi$ scattering data, consistent with dispersion relations. Namely, 
on a first stage one obtains 
simple fits to different, even conflicting, sets of data
for each partial wave up to 1.74 GeV, without any further constraint apart from unitarity. 
The resulting parameterizations 
form a set of simple ``Unconstrained Fits to Data''
that could be easily modified wave by wave in case new data would appear.
However, we check later that this set is 
not consistent with Forward Dispersion Relations
 up to 1.74 GeV.
Then, using this set as a starting point, 
one refines its parameters by imposing the dispersion 
relations without spoiling the data description. 
The resulting ``Constrained Fits to Data'' will be 
the main result of this work and
provide precise parameterizations describing the existing data, while 
being simultaneously consistent with Forward Dispersion relations up to 1.6 GeV as well as with three threshold sum rules. 
 Since these parameterizations are rather simple, we expect
that they will
become a useful tool for further studies, either theoretical and experimental, involving $\pi K$ scattering at some stage and particularly for the precise determination of resonance parameters. This was indeed the case of the parameterizations resulting from a similar analysis of $\pi\pi$ scattering.

\section{Kinematics and Notation}
\label{sec:notation}

As it is customary we will use the partial wave decomposition of the $\pi K$ scattering amplitudes 

\begin{equation}
T^I(s,t,u)=\frac{4}{\pi} \sum_l{(2l+1)P_l(\cos\theta) t^I_l(s)},
\label{ec:pwexpansion}
\end{equation}
where $s,t,u$ are 
the standard Mandelstam variables, satisfying $s+t+u=2(m_\pi^2+m_K^2)$ and
$\sigma(s)=2q_{K\pi}/\sqrt{s}$. 
The center of mass momentum of two particles with mass $m_1$ and $m_2$ is:
\begin{equation}
q_{12}(s)=\frac{1}{2\sqrt{s}}\sqrt{(s-(m_1+m_2)^2)(s-(m_1-m_2)^2)}.
\end{equation}
For later convenience we also define $\Sigma_{12}=m_1^2+m_2^2$ and $\Delta_{12}=m_1^2-m_2^2$.
Unless explicitly stated $m_1=m_K$ and $m_2=m_\pi$ and $q=q_{K\pi}$ in this work.
Note that we are working in the isospin limit of equal masses for all pions 
$m_\pi=139.57\,$MeV and equal masses for all kaons $m_K=496\,$MeV. 
We also use $m_\eta=547\,$MeV.

The elastic unitarity condition $\im t(s)=\sigma(s)\vert t(s)\vert^2$ implies that the elastic 
partial wave can be recast in terms of a real phase shift
\begin{equation}
t_l(s)=\frac{\hat t(s)}{\sigma(s)}=\frac{e^{i\delta_l(s)}\sin\delta_l(s)}{\sigma(s)}=
\frac{1}{\sigma(s)}\frac{1}{\cot\delta(s)-i},
\end{equation}
where we have introduced the ``Argand'' partial wave $\hat t (s)$ for later convenience.

In contrast, in the inelastic regime an inelasticity function is also introduced to write:
\begin{equation}
t_l(s)=\frac{\hat t(s)}{\sigma(s)}=\frac{\eta_l(s)e^{2i\delta_l(s)}-1}{2i\sigma(s)}.
\end{equation}

Later on we will also study the scattering at very low energies through the threshold parameters 
defined as:
\begin{equation}
{\rm Re}\,\hat{t}^I_l(s)\sim q^{2l+1}(a^I_l+b^I_lq^2+O(q^4)).
\label{eq:definitionab}
\end{equation}

Throughout this work we will also use the traditional spectroscopic notation,
naming the partial waves with isospin $I$ and 
angular momentum $l=0,1,2,3...$ as $S^I$, $P^I$, $D^I$ and $F^I-$waves..., respectively.

\section{Unconstrained Fits to Data}

\subsection{The Data}
\label{subsec:data}

Data on $\pi K$ scattering were obtained mostly during the 70's and the 80's,
measured indirectly from $KN\rightarrow K\pi N$ reactions
assuming they are dominated by the exchange of a single pion.

On the one hand, data on the $I=3/2$ $\pi K$-scattering cross sections 
was isolated in the early 70's using different reactions: Early experiments 
provided cross sections by studying $K^- d \rightarrow K^- \pi^- p p$ in Y. Cho et al. \cite{Cho:1970fb}, 
$K^- n \rightarrow K^- \pi^- p$ in A.M. Bakker et al. \cite{Bakker:1970wg} as well as 
$K^{\pm}p\rightarrow K^{\pm}\pi^{-}\Delta^{++}$
in B. Jongejans et al. \cite{Jongejans:1973pn}.
Since this $\pi K$ channel seems elastic up to at least 1.8 GeV, 
it is straightforward to obtain the phase shift.
Actually, this was done explicitly by D. Linglin et al. in \cite{Linglin:1973ci}
from their $K^- p\rightarrow K^- \pi^- \Delta^{++}$ analysis.
In general, the experiments in the earlier 70's have low statistics, which were improved
 by later experiments.
In particular, in 1977  P. Estabrooks et al. \cite{Estabrooks:1977xe}
performed a relatively high statistics analysis of $K^{\pm}p\rightarrow K^{\pm}\pi^{+}n$ 
and $K^{\pm}p\rightarrow K^{\pm}\pi^{-}\Delta^{++}$ at 13 GeV
to obtain the $I=3/2$ $\pi K$ component, also with no evidence of inelasticity up to 1.8 GeV
in $\pi K$ scattering. We will see that the differences between experiments are larger than
the statistical uncertainties they quote, which points at the existence of a sizable systematic uncertainty that we will have to estimate separately for each wave.

On the other hand, isospin $I=1/2$ scattering waves
have always been obtained in combination with those with $I=3/2$.
This was done for instance 
by R. Mercer et al. in \cite{Mercer:1971kn} using
$K^+ p \rightarrow K^+ \pi^- \Delta^{++}$ and $K^+ p \rightarrow K^0 \pi^0 \Delta^{++}$
reactions.  Due to low statistics, in order to separate different isospins, they
needed to combine their results  with the so-called ``World Data Summary Tape'',
an heterogeneous and not very precise collection of data that existed at that time.
As a consequence, the results for their $I=1/2$ and $3/2$ waves have huge uncertainties, which is why they
are usually neglected against later and more precise experiments.

As a matter of fact, what was really measured 
in scattering experiments was the $t_l=t^{1/2}_l+t^{3/2}_l/2$ combination. 
This was already studied with relatively high statistics in \cite{Estabrooks:1977xe} up to 1.85 GeV, 
but also in the experiment with the highest statistics so far that
was performed in the 80's by Aston et al. at the LASS Spectrometer \cite{Aston:1987ir}
at SLAC. This LASS experiment studied the $K^{-}p\rightarrow K^{-}\pi^{+}n$ reaction at 11 GeV
and obtained the same $\pi K$ partial wave combination up to 2.6 GeV.

The analysis needed to extract $\pi K$ scattering amplitudes from $KN\rightarrow K\pi N$ has 
several sources of systematic uncertainties, like
corrections to
the on-shell extrapolation of the exchanged pion or 
rescattering effects.
However, most experimental works only quote statistical uncertainties for each solution and for this reason conflicting data exist. This will be clearly seen in 
the figures below.
Thus, in our fits we sometimes add a systematic uncertainty
to different sets or to certain data points  which are in conflict with other data points in the same region. 
In the case of the most delicate and controversial wave, which is the $S^{1/2}$,
we have checked that the resulting data set and the fit are consistent with certain statistical tests explained in appendix~\ref{app:statistics}.

 In addition some ambiguities 
occur in the determination of the phase
that sometimes lead to different solutions for $\pi K$ 
scattering even within the same $KN\rightarrow K\pi N$ experiment. 
In the case of Aston et al. \cite{Aston:1987ir} these ambiguities appear above the region of interest for this work. In contrast, Estabrooks et al. \cite{Estabrooks:1977xe} do have four solutions above 1.5 GeV, but we only consider Solution B since it is the one qualitatively closer to Aston et al.

So far we have been discussing scattering data where
the $I=1/2$ state has always been obtained in combination 
with the $I=3/2$ one. However, it is also possible to obtain information
on $\pi K$ scattering from the decays of heavier particles.
In particular, when $\pi K$ are the only strongly interacting particles 
in the decay, Watson theorem implies that, in the $\pi K$ elastic region,
the phase of the global process should be the same as the scattering phase shift.
In particular, the phase-shift difference between S and P waves with $I=1/2$
have been measured from $D^+\rightarrow K^-\pi^+e^+\nu_e$ by the BaBar Collaboration \cite{delAmoSanchez:2010fd} and recently by the BESIII Collaboration \cite{Ablikim:2015mjo}.
 The results are 
very consistent with the LASS experiment, but their uncertainties are too large and will not be included in our fits, although we will show them for completeness.

Moreover, there are measurements of the $I=1/2$ phase of the 
$K \pi$ S-wave amplitude obtained from Dalitz plot analyses 
of $D^+\rightarrow K^-\pi^+\pi^+$ by the E791 \cite{Aitala:2005yh}, FOCUS \cite{Pennington:2007se} and CLEO-c \cite{Bonvicini:2008jw} collaborations, as well as a recent similar analysis 
of $\eta_c\rightarrow K\bar K \pi$ by the BaBar Collaboration \cite{Lees:2015zzr}. These phases (and amplitudes) are not necessarily those of $\pi K$ scattering due to the presence of a third strongly interacting particle, which invalidates the use of Watson's Theorem. 
However, a posteriori comparison with
the scattering data has shown that, within the large uncertainties and
at least in the elastic region, 
the resulting phase (but not the amplitude) is very similar to that of LASS.
This means that the effect of the third particle on the phase is rather constant and almost amounts to a global shift. But these data cannot really be interpreted as a scattering phase beyond this qualitative agreement and are therefore not included in our fits. Nevertheless we will show and discuss them in comparison with our results.

\subsection{General form of our parameterizations}

Each partial wave will now be fitted to 
the existing data up to $\sim$1.7 GeV, 
which means that we will only fit $S$, $P$, $D$ and $F$ waves,
since there are no data for $G$, $H$ and higher waves below 1.8 GeV.
In this first stage, the fit to a wave with a given angular momentum will be performed 
independently of other waves with different angular momentum, 
by means of simple functions, without imposing any dispersive constraint.
For this reason the resulting set of partial waves will be 
called ``Unconstrained Fit to Data'' (UFD).
When possible, as in waves which are elastic in the whole energy range, a single functional form will be used throughout the whole energy region. However, for more complicated waves different functional forms
will be used in different regions. Typically these piecewise functions will be matched at thresholds demanding continuity. 

We would like to add a word of caution here. The data are not precise nor numerous enough to exclude 
large fluctuations between successive data points, particularly in certain energy regions. One could devise complicated parameterizations that would pass through every single data point, or even produce fluctuations between points. In this work we are assuming that such fluctuations do not occur and that the data can be correctly fitted with simple and relatively smooth parameterizations. The size of the uncertainties thus depends on this assumption. 
 The parameterizations we describe below are the ones we have finally chosen because they satisfy the above assumption and yield uncertainty bands which do not show wild fluctuations or become too large in a region where the data spread does not require so. In particular, we have explored different kinds of conformal parameterizations (with different centers and more terms in the expansion, see Appendix \ref{app:conformal}), we have tried simple polynomials in different variables, including orthogonal
polynomials in a given region, adding or removing resonant shapes, etc.  Since all them fit the data, their central result is not too different from our final choice. Except in some few relevant cases, we spare the reader from explaining the caveats that affect these many other parameterizations we tried. We just present below our final choice. Moreover, for a given parameterization, and once the systematic uncertainty that affects the data has been estimated, we decide to stop adding parameters 
when the $\chi^2/dof$ is close or less than one. Of course, the size of the final uncertainties depend on our educated guess of systematic uncertainties, which, as we will see, dominate the final error bands in many cases.

\subsubsection{Partial waves in elastic regions}
\label{subsec:pwelastic}

For the elastic regions, in which a partial wave can be recast in terms of just a phase-shift,
we will use a conformal expansion of the type:
\begin{equation}
\cot\delta_l(s)=\frac{\sqrt{s}}{2q^{2l+1}}F(s)\sum_n{B_n \omega(s)^n},
\label{eq:generalconformal}
\end{equation}
where $F(s)=1$ except for scalar waves that have an Adler zero at $s_{Adler}$, 
in which case $F(s)=1/(s-s_{Adler})$, or for waves that exhibit a clear narrow resonance
and whose phase shift crosses $\pi/2$ at $m_r$, in which case $F(s)=(s-m_r^2)$.
In addition, the conformal variable is defined as:
\begin{equation}
\omega(y)=\frac{\sqrt{y}-\alpha \sqrt{y_0-y}}{\sqrt{y}+\alpha \sqrt{y_0-y}}, \quad y(s)=\left(\frac{s-\Delta_{K\pi}}{s+\Delta_{K\pi}}\right)^2.
\label{eq:conformalvars}
\end{equation}
This change of variables, which maps the complex $s$ plane into the unit circle, 
is relatively similar to those used for $\pi\pi$ scattering in \cite{Pelaez:2004vs,Caprini:2008fc}
or $\pi K$ scattering in \cite{Cherry:2000ut}, and is explained in detail in Appendix \ref{app:conformal}.
Suffices here to say that, by taking full advantage of the analytic properties of the partial waves
in the complex plane, such a conformal expansion ensures a rapid convergence of the series and 
no more than three $B_i$ coefficients are needed
for the fits to each wave in the elastic region.
The $y_0\equiv y(s_0)$ and $\alpha$ constants are fixed, not fitted, 
for each partial wave. The $s_0$ parameter sets the maximum energy at which 
this mapping is applicable on the real axis, whereas $\alpha$ 
fixes the energy where the expansion is centered.

\subsubsection{Partial waves in Inelastic regions}
\label{subsec:pwinelastic}

The parameterizations of partial waves in the inelastic region
have to accommodate several resonant structures 
that have been observed and ensure a continuous matching with 
the elastic parameterization.
Note that for the $D^{1/2}$ and $F^{1/2}$-waves, since data only exist
in the inelastic region, we will use a unified inelastic formalism
in the whole energy region, which reduces 
to the elastic case below $K\eta$ threshold.

We have tried different parameterizations, 
like  polynomial fits in powers of the 
the $\pi K$, $K\eta$ momenta, or the $s$ or $\sqrt{s}$ variables. However 
such fits tend to have small uncertainties close to the elastic region and very 
large as the energy increases, which does not necessarily reproduce the uncertainty observed in the data and leads to huge correlations. 
As other authors before \cite{Buettiker:2003pp}, we have found that 
it is more efficient to describe this region with products 
of exponential or rational functions, which are more flexible 
to accommodate resonant structures and whose resulting 
uncertainty bands are more uniform throughout the fit region. 
Moreover, the use of products of functions allows for 
a straightforward implementation of unitarity, which is done as follows
\begin{equation}
t_l(s)=\frac{1}{2i\sigma(s)}\left( \prod_n{S_n}(s)-1\right).
\label{eq:inelunit}
\end{equation}
The $S_n$ could either have the form of a non-resonant background
\begin{equation}
S_n=S_{n}^b=\exp\left[2i q_{ij}^{2l+1}(\phi_0+\phi_1 q_{ij}^2+...)\right],
\label{eq:inelback}
\end{equation}
with $\phi_k$ real parameters, or a resonant-like form
\begin{equation}
S_n=S_n^r=\frac{s_{rn}-s+i(P_{n}(s)-Q_{n}(s))}{s_{rn}-s-i(P_{n}(s)+Q_{n}(s))},
\label{eq:inelres}
\end{equation}
where $s_{rn}$ are real parameters and $P_{n}(s)$ and $Q_{n}(s)$ are polynomials that have the same sign over the inelastic region. 
Using the equations above, $\vert S_n\vert \leq1$ and inelastic unitarity is satisfied.
If these polynomials were constant, one would recover the simplest Breit-Wigner 
formula for $S_n^r$.
We will explain in the following subsections the choice of polynomials for different waves.
Continuity with the elastic region is imposed by fixing the $P_{n}(s)$ polynomial 
of the $S_n$ that has a pole with the lowest $s_{rn}$.
This formalism is a modification of the parameterizations 
used in \cite{Buettiker:2003pp} for the high energy region
\footnote{We thank B. Moussallam for kindly providing us with his parameterizations.}. When reducing this parameterization to the elastic case
$Q_n$ is set to zero, which as commented above is of relevance for the $D^{1/2}$ and $F^{1/2}$ waves.

Note that close to a resonance, 
each of the $S_n^r$ functions bear some resemblance to a Breit-Wigner form,
but the actual parameters of a resonance have to be calculated with the full partial wave and not obtained from an individual $S_n^r$. Let us remark once again that
when combining the $S_n$ in the complete functional form of $t_l$, 
unitarity has been enforced exactly. This
would not occur in a simple sum of Breit-Wigner amplitudes, which would violate unitarity.

We will use partial waves to describe data up to $\sim 1.7\,$GeV.
Beyond that energy we will use Regge theory to describe 
the amplitudes, as we will see in Sect.\ref{sec:Regge} below.

\subsection{S-waves}

\subsubsection{I=3/2 S-wave}
Let us then start by describing our simple fit to the $I=3/2$ S-wave, since 
data for this wave exist independently of other waves.
Once again, we emphasize that there is no evidence so far of inelasticity up to 
$\sim$1.8 GeV, and thus we will consider this wave
as elastic up to that energy. Hence, as commented above 
and explained in more detail in the appendix, 
 we will use the following simple 
functional form to describe the phase shift:
\begin{equation}
\cot\delta_0^{3/2}(s)=\frac{\sqrt{s}}{2q(s_{Adler}-s)}(B_0+B_1\omega(s)+B_2\omega(s)^2).
\label{eq:cot32}
\end{equation}
Note we have explicitly factorized the Adler zero, which we will set to 
its leading order within Chiral Perturbation Theory, i.e., $s_{Adler}=\Sigma_{K\pi}$. 
For this wave, the constants that define the conformal variable $\omega$ in Eq.\eqref{eq:conformalvars}
are fixed to 
\begin{eqnarray}
\alpha=1.4, \quad s_0=(1.84 \, {\rm GeV})^2.
\end{eqnarray}

\begin{figure}
\centering
\includegraphics[scale=0.33]{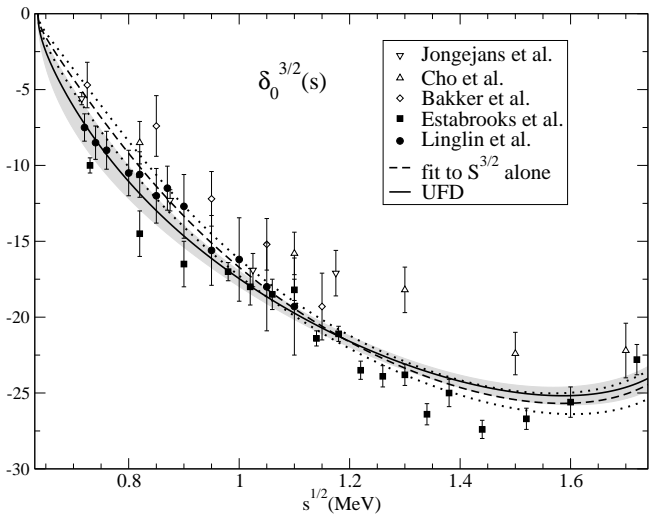}
 \caption{\rm \label{fig:S32} 
Experimental data on the $S^{3/2}$ phase shift, $\delta_0^{3/2}(s)$.
The data come from \cite{Cho:1970fb} (Y. Cho et al.),  \cite{Bakker:1970wg} (A.M. Bakker et al.),
\cite{Jongejans:1973pn} (B. Jongejans et al.),  \cite{Linglin:1973ci} (D. Linglin et al.)
and \cite{Estabrooks:1977xe} (P. Estabrooks et al.). The dashed line shows our fit to these data and the dotted lines enclose its uncertainty band.
The continuous line represents our unconstrained fit including also the 
data on $t^{1/2}_0+t^{3/2}_0/2$, whose uncertainty is represented by the gray band.}
\end{figure}

The existing data are shown in Fig.\ref{fig:S32}. 
There is a relatively fair agreement between different experiments below 1.1 GeV.
However we can already notice
some incompatible points between the Bakker et al. \cite{Bakker:1970wg} and Estabrooks et al. \cite{Estabrooks:1977xe} data sets, 
mostly due to the very small uncertainty of some points in the latter set. 
Note also the large variations between the uncertainties of successive data points
in the Estabrooks et al. set.
Above 1.1 GeV the two data sets that exist are largely incompatible.
It is clear that some systematic uncertainty exists. 

Therefore, we have fitted the data in Fig.\ref{fig:S32} in two ways, 
either adding a constant systematic uncertainty of $1^{\rm o}$ or multiplying 
the existing statistical uncertainties by a factor of 2, which is chosen 
so that  the resulting $\chi^2/d.o.f.$ is slightly less than one. 
The resulting fits are rather similar,
but we have preferred the uncertainty band of the first because the systematic uncertainty is not correlated to the statistical one.
In addition, the second approach satisfies much worse the 
threshold sum rules that we will check in the next sections.
The result of our fit, with the estimate of
systematic uncertainty added 
to the statistical uncertainties, is $\chi^2/d.o.f.=37/(44-3+1)$.

\begin{table}[h] 
\caption{Parameters of the S$^{3/2}$-wave.} 
\centering 
\begin{tabular}{c c c } 
\hline\hline  
\rule[-0.05cm]{0cm}{.35cm}Parameter & UFD & CFD \\ 
\hline 
\rule[-0.05cm]{0cm}{.35cm}$B_0$ & 2.25 $\pm$0.04  & 2.27 $\pm$0.04\\
\rule[-0.05cm]{0cm}{.35cm}$B_1$ & 4.21 $\pm$0.17  & 3.94 $\pm$0.17\\ 
\rule[-0.05cm]{0cm}{.35cm}$B_2$ & 2.45 $\pm$0.50  & 3.36 $\pm$0.50\\ 
\hline 
\end{tabular} 
\label{tab:S32pa} 
\end{table}

Had we considered only two $B_k$ parameters, 
the fit would yield an 80\% larger $\chi^2/d.o.f.$,
whereas with four it would decrease 
by 15\%. Since three parameters as in Eq.\ref{eq:cot32}
already provide a  $\chi^2/d.o.f.<1$
we do not consider necessary to have a fourth parameter.
We show this 
fit as a dashed line in Fig.\ref{fig:S32}, where the uncertainty band
is delimited by the dotted lines. 

Still this is not our final fit 
because there is also 
experimental information on the $t_S \equiv t^{1/2}_0+t^{3/2}_0/2$ combination.
In the next subsection we will explain how the fit to
the $t_S$ data produces a small modification on 
the $S^{3/2}$-wave.
The result provides the final $S^{3/2}$ parameterization, which is also shown
in Fig.\ref{fig:S32} as a thick continuous line whose uncertainties are covered by the gray band. Since no dispersion relation has been imposed yet, 
this result will be called Unconstrained Fit to Data (UFD), whose
parameters are found in Table~\ref{tab:S32pa}.
The Constrained Fit to Data (CFD) in that table will be discussed 
later in Sec.\ref{sec:CFD}.
In the Figure it can be noticed that this UFD result is similar to the fit
to the $S^{3/2}$-wave data alone that has been described in this subsection.

\subsubsection{I=1/2 S-wave}

For this wave, inelasticity 
has been measured above 1.3 GeV and for the most 
part it is due to the $K \eta$ state rather than to states with more than two mesons.
Hence, we are going to parameterize the amplitude using the
elastic formalism of Subsec.\ref{subsec:pwelastic} below $K \eta$ threshold, 
and with the inelastic formalism of Subsec.\ref{subsec:pwinelastic} 
above that threshold.

Thus, for $(m_K+m_\pi)^2\leq s \leq (m_K+m_\eta)^2$
we will use a conformal expansion 
of the type in Eq.\eqref{eq:generalconformal},
namely:
\begin{equation}
\cot\delta_0^{1/2}(s)=\frac{\sqrt{s}}{2q(s-s_{Adler})}(B_0+B_1\omega).
\label{eq:cot12}
\end{equation}
Once again we have explicitly factorized the Adler zero, which we have set to 
its leading order within Chiral Perturbation Theory value:
\begin{equation}
s_{Adler}=\Big(\Sigma_{K\pi}+2\sqrt{\Delta_{K\pi}^2+m_K^2m_\pi^2}\,\Big)/5\simeq 0.236\,{\rm GeV}^2.
\end{equation}
As explained in Appendix \ref{app:conformal}, for this wave 
it is convenient to fix the constants that define the center of the conformal variable $\omega$ in Eq.\eqref{eq:conformalvars} to the following values
\begin{eqnarray}
\alpha=1.15, \quad s_0=(1.1 \,{\rm GeV})^2.
\end{eqnarray}

The parameters obtained for the best Unconstrained Fit to Data (UFD) 
are given in the first column of Table~\ref{tab:Selparam}.

\begin{table}[h] 
\caption{Parameters of the elastic S$^{1/2}$-wave.} 
\centering 
\begin{tabular}{c c c } 
\hline\hline  
\rule[-0.05cm]{0cm}{.35cm}Parameter & UFD & CFD \\ 
\hline
\rule[-0.05cm]{0cm}{.35cm}$B_0$ & 0.411  $\pm$0.007        & 0.411  $\pm$0.007\\
\rule[-0.05cm]{0cm}{.35cm}$B_1$ & 0.181 $\pm$0.034         & 0.162 $\pm$0.034\\ 
\hline 
\end{tabular} 
\label{tab:Selparam} 
\end{table}

In contrast, in the $s\geq(m_K+m_\eta)^2$ region we will implement the inelastic formalism of Eqs.\eqref{eq:inelunit},\eqref{eq:inelback},\eqref{eq:inelres}
as follows:
\begin{equation}
t_0^{1/2}(s)=\frac{S_0^b S_1^r S_2^r-1}{2i\sigma(s)},
\end{equation} 
where
\begin{equation}
S_0^b=\exp[2iq_{\eta K}(\phi_0+\phi_1 q_{\eta K}^2)].
\end{equation}
For $S^r_1$ we use Eq.\eqref{eq:inelres} with
\begin{eqnarray}
P_1(s)&=&(s_{r1}-s)\beta+e_1G_1 \frac{p_1(q_{\pi K})}{p_1(q_{\pi K}^r)}\frac{q_{\pi K}-\hat{q}_{\pi K}}{q_{\pi K}^r-\hat{q}_{\pi K}},\\
Q_1(s)&=&(1-e_1)G_1 \frac{p_1(q_{\pi K})}{p_1(q_{\pi K}^r)}\frac{q_{\eta K}}{q_{\eta K}^r}\Theta_{\eta K}(s),
\end{eqnarray}
where $p_1(x)=1+a x^2+b x^4$, $q_{i j}^r=q_{i j}(s_{r})$, $\hat{q}_{i j}=q_{i j}((m_{\eta}+m_K)^2)$
and $\Theta_{\eta K}(s)=\Theta(s-(m_K+m_\eta)^2)$ is the step function at the $K\eta$ threshold.
In addition, for $S^r_2$ we use Eq.\eqref{eq:inelres} with
\begin{eqnarray}
P_2(s)&=&e_2G_2 \frac{p_2(q_{\pi K})}{p_2(q_{\pi K}^r)}\frac{q_{\pi K}-\hat{q}_{\pi K}}{q_{\pi K}^r-\hat{q}_{\pi K}},\\
Q_2(s)&=&(1-e_2)G_2 \frac{p_2(q_{\pi K})}{p_2(q_{\pi K}^r)}\frac{q_{\eta K}}{q_{\eta K}^r}\Theta_{\eta K}(s),
\end{eqnarray}
with $p_2(x)=1+c x^2$.

By matching the elastic and inelastic parameterizations at 
the $K \eta$ threshold
we only need to demand continuity, which is ensured by defining 
$\beta\equiv 1/\cot\delta_0^{1/2}((m_K+m_\eta)^2)$,
where $\delta_0^{1/2}$ is calculated here with the elastic parameterization in Eq.\eqref{eq:cot12}.

\subsubsection{$t_S$-wave}

Nevertheless, as already explained above, we do not fit the $S^{1/2}$-wave alone,
but in the $t_S\equiv t^{1/2}_0+t^{3/2}_0/2 $ combination that was originally measured.
Let us then define
\begin{equation}
t_S (s)=\vert t_S(s)\vert e^{i\phi_S(s)},
\end{equation}
and remark that, since $\vert t_S\vert$ and $\phi_S$ were measured independently,
we will fit them both.  
In order to compare with data is convenient to use the normalization
\begin{equation}
\hat t_S (s)=t_S(s) \sigma(s).
\end{equation}

\begin{figure*}
 \centering
 \includegraphics[scale=0.66]{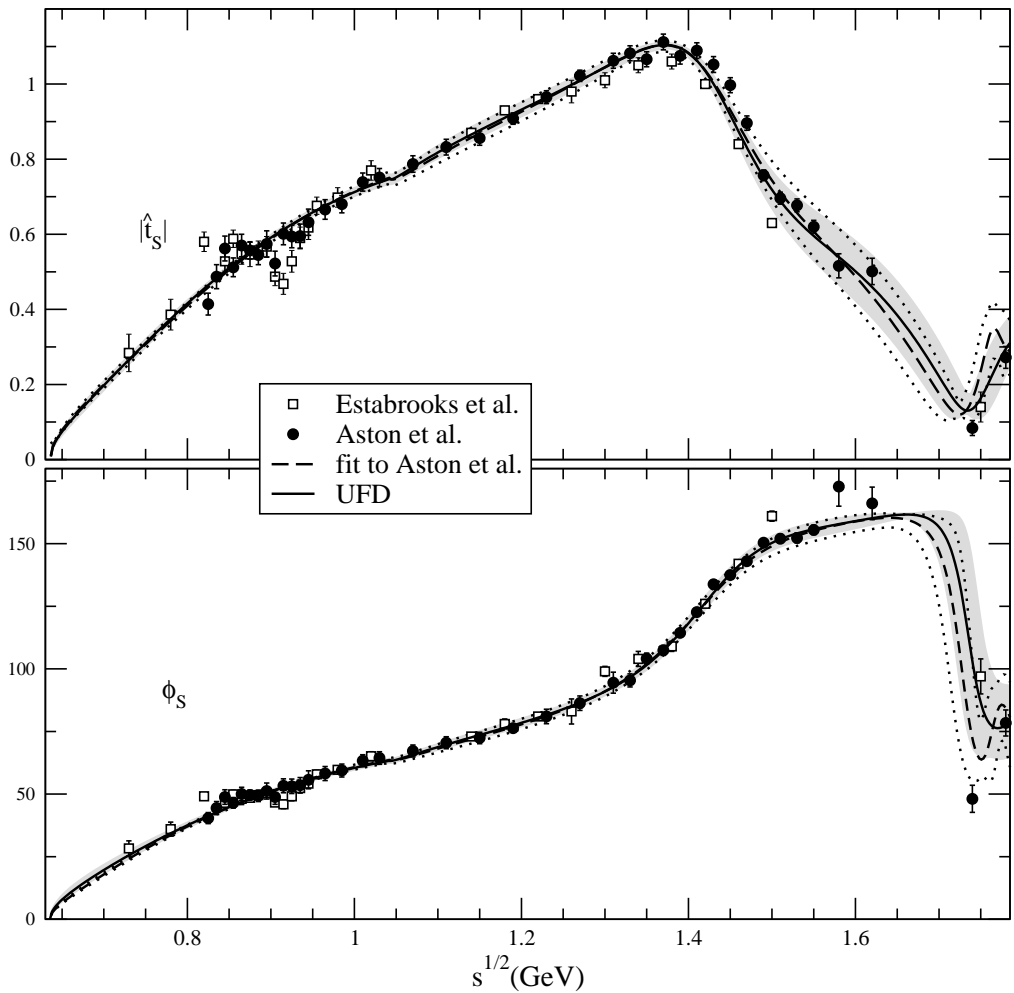}
%\vspace{-.5cm}
  \caption{\rm \label{fig:Sampli} 
Data on $\hat t_S(s)$ from Estabrooks et al. \cite{Estabrooks:1977xe} and Aston et al. \cite{Aston:1987ir}.
The upper panel shows $\vert \hat t_S(s)\vert$ whereas the lower
one shows $\phi_S(s)$, which were measured independently.
The continuous line is our unconstrained fit (UFD), 
whose uncertainties are covered by the gray band. 
For comparison we show, as a dashed line, a fit only to the data in  
\cite{Aston:1987ir}, whose corresponding uncertainties
are delimited by the dotted lines. 
}
 \end{figure*}

Thus, in Fig.\ref{fig:Sampli} we show the data on $\hat t_S$ 
and the result of our Unconstrained Fit Data (UFD).
The upper panel shows $\vert \hat t_S\vert$ whereas the lower one shows $\phi_S$.
The combined $\chi^2/d.o.f.$ of the $S^{1/2}$ and $S^{3/2}$ data fits
is $\chi^2/d.o.f.=168/(182-15+1)$ with the UFD parameters
provided in Table~ \ref{tab:Sinparam}.
The $e_1$ parameter was initially left free but 
it comes out practically indistinguishable from 1, with tiny uncertainties, and has been later fixed to 1 for practical purposes.

\begin{table}[h] 
\caption{Parameters of the S$^{1/2}$ inelastic fit.} 
\centering 
\begin{tabular}{c c c } 
\hline\hline  
\rule[-0.05cm]{0cm}{.35cm}Parameters & UFD & CFD \\ 
\hline 
\rule[-0.05cm]{0cm}{.35cm}$\phi_0$ & -0.20       $\pm$0.04GeV$^{-1}$   & -0.19       $\pm$0.04GeV$^{-1}$\\
\rule[-0.05cm]{0cm}{.35cm}$\phi_1$ & 4.76         $\pm$0.25GeV$^{-3}$  & 5.03         $\pm$0.25GeV$^{-3}$\\ 
\rule[-0.05cm]{0cm}{.35cm}$a$ & -5.22            $\pm$0.04GeV$^{-2}$   & -5.20            $\pm$0.04GeV$^{-2}$\\ 
\rule[-0.05cm]{0cm}{.35cm}$b$ &  7.57             $\pm$0.13GeV$^{-4}$  & 7.60             $\pm$0.13GeV$^{-4}$\\ 
\rule[-0.05cm]{0cm}{.35cm}$c$ & -1.72             $\pm$0.07GeV$^{-2}$  & -1.73             $\pm$0.07GeV$^{-2}$\\ 
\rule[-0.05cm]{0cm}{.35cm}$\sqrt{s_{r1}}$ & 1.399 $\pm$0.006GeV    & 1.401 $\pm$0.006GeV\\ 
\rule[-0.05cm]{0cm}{.35cm}$\sqrt{s_{r2}}$ & 1.815 $\pm$0.017GeV    & 1.817 $\pm$0.017GeV\\ 
\rule[-0.05cm]{0cm}{.35cm}$e_1$ & 1                                  & 1\\ 
\rule[-0.05cm]{0cm}{.35cm}$e_2$ & 0.184          $\pm$0.033         & 0.183           $\pm$0.033\\ 
\rule[-0.05cm]{0cm}{.35cm}$G_1$ & 0.499           $\pm$0.017GeV    & 0.497           $\pm$0.017GeV\\ 
\rule[-0.05cm]{0cm}{.35cm}$G_2$ & 0.29            $\pm$0.12GeV     & 0.28            $\pm$0.12GeV\\
\hline 
\end{tabular} 
\label{tab:Sinparam} 
\end{table}

From Fig.\ref{fig:Sampli} it can be easily noted that there are data 
points which are largely incompatible with one another, not only between the two different experiments
\cite{Estabrooks:1977xe,Aston:1987ir}, but even among the successive data points of Estabrooks et al. \cite{Estabrooks:1977xe}. Thus, it seems clear that there are some systematic errors not 
covered by the 
experimental uncertainties. Since these are the most controversial waves, here we have followed a more elaborated procedure to estimate the uncertainties of the resulting fit. In particular, 
we follow one of the techniques suggested in \cite{Perez:2015pea}, 
which, in brief, consists of running Gaussianity tests on the data with respect to the fit
and enlarge the uncertainties of those data points that spoil the test. This yields a new fit upon which
the procedure is iterated until the Gaussianity test is satisfied. The details of this method are given in Appendix \ref{app:statistics}.
We show in Fig.\ref{fig:Sampli}, as a gray band,
the resulting uncertainty of our fit to those data and the $I=3/2$ data already discussed in the previous section.

\begin{figure}
\centering
\includegraphics[scale=0.33]{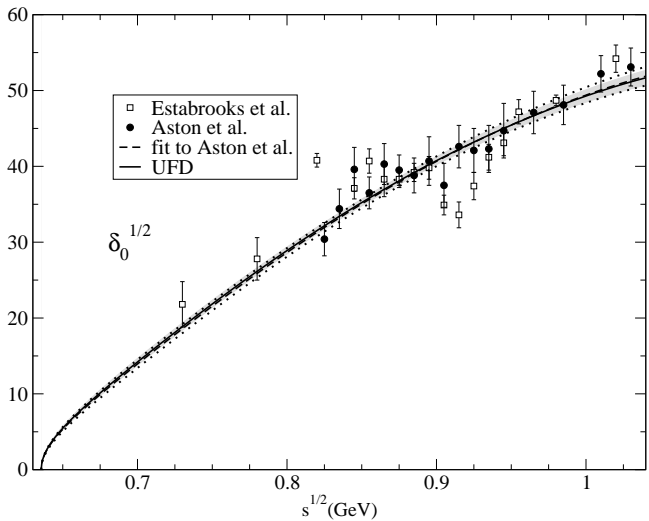}
 \caption{\rm \label{fig:swavephaseshift}
$S^{1/2}$-wave phase shift below $K\eta$ threshold. In this region  
the amplitude is elastic in practice. The continuous line is our UFD result
whose uncertainty is covered by the gray band.
Data points extracted from 
Estabrooks et al. \cite{Estabrooks:1977xe} and Aston et al.
 \cite{Aston:1987ir}.
As explained in the text, we do not fit this wave individually, but in combination with the $I=3/2$, as it was originally measured.
The dashed curve is the fit to Aston et al. \cite{Aston:1987ir} data alone
 and the dotted lines cover the corresponding uncertainty band.}
\label{Fig:Swave12}
\end{figure}

In the literature it is rather usual \cite{Buettiker:2003pp, Bugg:2009uk, Zhou:2006wm, Ishida:1997wn} 
to neglect the Estabrooks et al. data, although it is not always the case \cite{Jamin:2000wn}.
To be able to compare with this choice, we have thus
considered a fit to the $I=3/2$ data together with 
only the $I=1/2$ data set of Aston et al.\cite{Aston:1987ir}, which is much smoother than that of 
Estabrooks et al. \cite{Estabrooks:1977xe}, 
particularly below 1.5 GeV. 
In this case we have not added any systematic uncertainty 
and the result is shown in Fig.\ref{fig:Sampli} as a dashed line, 
which almost overlaps with our previous fit up to 1.5 GeV, and has a very similar uncertainty band
represented as the area between dotted lines. However, above 1.5 GeV 
and up to 1.7 GeV, the Aston et al.\cite{Aston:1987ir}
set is not so consistent. For instance,
it is well known 
that two of its points violate unitarity \cite{Jamin:2000wn} (which we have always removed from our fits).
Nevertheless, it is still compatible with our previous fit within uncertainties. 
Since here we want to stay on the conservative side, 
we have decided
not to neglect the Estabrooks et al. data. Thus, 
from now on we will consider our UFD result only, which describes both sets. We will repeat this comparative exercise for other waves, but in all cases we will keep the UFD result obtained by fitting both sets when they exist.

With the combined fit to the $I=1/2$ and $I=3/2$ data we can separate the 
results for each isospin partial wave. The UFD 
result for the $I=3/2$ S-wave was already shown in its elastic region in Fig.\ref{fig:S32},
whereas we show now in Fig.\ref{Fig:Swave12} the resulting $I=1/2$ S-wave phase shift. Note once again 
that in the elastic region our UFD result is almost identical to the fit to Aston et al. data alone.

\subsubsection{S-wave scattering Lengths}

Once we have fitted the data on the two S-waves, we can  use our UFD 
parameterizations to
obtain the scattering lengths 
defined in Eq.\eqref{eq:definitionab}, which we show  in Table~\ref{tab:slengths}. Note  they are in fair agreement with
other existing values in the literature, also provided in the table.

\begin{table}[h] 
\centering 
\caption{S-wave scattering lengths from our UFD and CFD sets, in $m_\pi^{-1}$ units, compared to other values in the literature.
\label{tab:slengths} }
\begin{tabular}{l r r} 
\hline
\rule[-0.15cm]{0cm}{.55cm}  
& $m_\pi a_0^{1/2}$ &$m_\pi a_0^{3/2}$\\ 
\hline\hline
\rule[-0.15cm]{0cm}{.55cm}   B{\"u}ttiker et al. Ref.\cite{Buettiker:2003pp}   & 0.224$\pm$0.022   & -0.0448$\pm$0.0077 \\ 
\rule [-0.15cm]{0cm}{.55cm} Dobado \& Pel\'aez  Ref.\cite{Dobado:1996ps}      & 0.155$\pm$0.012 &$-0.049\pm0.004$\\
\rule[-0.15cm]{0cm}{.55cm}   Jamin et al. Ref.\cite{Jamin:2000wn}       & 0.18 & -0.12  \\
\rule[-0.15cm]{0cm}{.55cm}   Bugg Ref.\cite{Bugg:2009uk}        & 0.195$\pm$0.006 & -\\
\rule[-0.15cm]{0cm}{.55cm}   Zhou \& Zheng Ref.\cite{Zhou:2006wm}        & 0.219$\pm$0.034  & -0.042$\pm$0.002 \\ 
%\rule[-0.15cm]{0cm}{.55cm}   Ref.\cite{Ishida:1997wn}      &   &  \\ 
\hline \hline 
\rule[-0.15cm]{0cm}{.55cm}   UFD, this work                & 0.22$\pm$0.01 & $-0.10^{+0.03}_{-0.05}$  \\
\rule[-0.15cm]{0cm}{.55cm}  {\bf CFD, this work    }            & {\bf 0.22$\pm$0.01} & {\bf -0.054$^{+0.010}_{-0.014}$} \\\hline
\end{tabular} 
\end{table}

There is a renewed interest on these quantities 
due to recent lattice calculations \cite{Beane:2006gj} and also due to the
experimental measurement
by the DIRAC collaboration \cite{Adeva:2014xtx}
\begin{equation}
\frac{1}{3}\left(a_0^{1/2}-a_0^{3/2}\right)=0.11^{+0.09}_{-0.04} \, m_\pi^{-1},\quad ({\rm DIRAC})
\label{eq:dirac}
\end{equation}
which was not determined from scattering experiments, but
from the formation of $ \pi K$ atoms. From our UFD set we find:
\begin{equation}
\frac{1}{3}\left(a_0^{1/2}-a_0^{3/2}\right)=0.108^{+0.018}_{-0.010} \, m_{\pi}^{-1}.\quad ({\rm UFD})
\label{ec:DIRACUFD}
\end{equation}
 Note that our  uncertainties are  smaller, 
by roughly an order of magnitude, than the present direct experimental knowledge. 
We have explicitly checked that including or not the DIRAC value does not change the result of our fits.

\subsection{$P$-waves}

\subsubsection{$I={3/2}$ P-wave}
\label{subsec:P32UFD}

\begin{figure}
\centering
\includegraphics[scale=0.33]{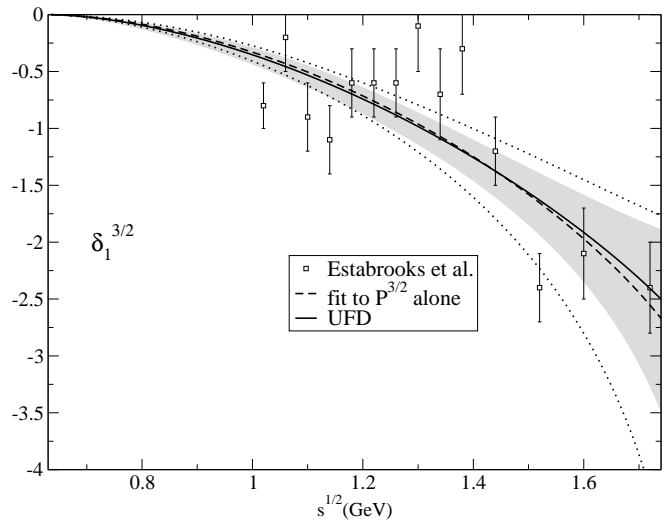}
 \caption{\rm \label{fig:p32wavephase} 
Data on the $I=3/2$ $P$-wave from Estabrooks et al. \cite{Estabrooks:1977xe}.
We also show our UFD result as solid line with a gray uncertainty band, 
which is obtained by fitting these data 
together with the data on the $t_P=t^{1/2}_1+t^{3/2}_1/2$ combination.
For comparison we show with a dashed line a fit only 
to the data in this figure, whose uncertainty is 
delimited by the dotted lines. 
}
\end{figure}

Only Estabrooks et al.\cite{Estabrooks:1977xe} provide data for the $I={3/2}$ P-wave
phase-shift up to 1.74 GeV, which we show in Fig.\ref{fig:p32wavephase}. 
As it can be noticed in the figure,
this wave is rather small. Namely, below 1.1 GeV its phase shift is less 
than $1^\degree$, below 1.4 GeV is less than $2^\degree$ and below 1.74 GeV it is less than $3^\degree$.
There is no inelasticity measured up to 1.74 GeV so that we 
will parameterize this partial wave with a conformal expansion as in Eqs.\eqref{eq:generalconformal} and \eqref{eq:conformalvars}:
\begin{equation}
\cot\delta_1^{3/2}(s)=\frac{\sqrt{s}}{2q^3}(B_0+B_1\omega).
\end{equation}
Let us remark that the $\alpha$ parameter
that defines the conformal variable $\omega$ (see Eq.\eqref{eq:conformalvars}) has been chosen
so that the center of the conformal expansion lies on the center of the region where data exists. Thus, for this wave we have set
\begin{eqnarray}
\alpha=1.45, \quad s_0=(1.84 \,{\rm GeV})^2.
\end{eqnarray}

The existence of systematic uncertainties
is evident from Fig.\ref{fig:p32wavephase}.
If we make a naive fit without taking these systematic effects into account the resulting $\chi^2/d.o.f.\simeq 2$.
Hence, we have included an estimation of the systematic uncertainty in our fits
by multiplying the data statistical uncertainties by $\sqrt{2}$. Two conformal parameters are enough to describe this wave and no significant improvement is obtained by considering a third one.

As it happened with the $S^{3/2}$ wave, our final fit for the $I={3/2}$ P-wave is obtained
by fitting simultaneously the data for this wave alone together with the
data on the $t_P \equiv t^{1/2}_1+t^{3/2}_1/2$ combination obtained
by Estabrooks et al. \cite{Estabrooks:1977xe} and Aston et al. \cite{Aston:1987ir}.
The resulting unconstrained fit to data (UFD) is shown in Fig.\ref{fig:p32wavephase}, 
where the gray band covers its uncertainty. 
The corresponding UFD parameters are listed in Table~\ref{tab:P32param}.

\begin{table}[h] 
\caption{Parameters of the P$^{3/2}$-wave.} 
\centering 
\begin{tabular}{c c c } 
\hline\hline  
\rule[-0.05cm]{0cm}{.35cm}Parameter & UFD & CFD \\ 
\hline 
\rule[-0.05cm]{0cm}{.35cm}$B_0$ & -14.8 $\pm$2.6  & -15.6 $\pm$2.6\\
\rule[-0.05cm]{0cm}{.35cm}$B_1$ & 2.7   $\pm$7.4  & -2.2   $\pm$7.4\\ 
\hline 
\end{tabular} 
\label{tab:P32param} 
\end{table}

Also in Figure~\ref{fig:p32wavephase} we show, as a dashed line, the result when fitting only the data on that wave.
Its corresponding uncertainty band is delimited by dotted lines. As we can see it is almost indistinguishable from our UFD result, for which we have also fitted
data on $t_P \equiv t^{1/2}_1+t^{3/2}_1/2$, as we will see next.

\subsubsection{$I={1/2}$ P-wave}

The $I$=1/2-wave is only measured in scattering experiments together
with the $I$=3/2-wave in the $t_P$ combination defined just above. 
Although in the literature it is frequent to 
neglect the $P^{3/2}$-wave, because as we have just seen is very small, 
we will keep it in our fits for completeness.

Let us then discuss the $P^{1/2}$-wave
in the elastic region, i.e. $s\leq{(m_{\eta}+m_K)^2}$,
for which  we use a conformal fit to describe the data, namely,
\begin{equation}
\cot\delta_1^{1/2}(s)=\frac{\sqrt{s}}{2q^3}(m_r^2-s)(B_0+B_1\omega+B_2\omega^2).
\label{eq:cot121}
\end{equation}
Note we have explicitly extracted an $(m_r^2-s)$ factor
so that the phase crosses $\pi/2$ at the energy of the 
peak associated to the $K^*(892)$ resonance, which is the dominant feature of this wave in the elastic region.
As explained in Appendix \ref{app:conformal}, the $\alpha$ and $s_0$ parameters, which define the conformal variable $\omega$ in Eq.\eqref{eq:conformalvars},
are fixed from the choice of the center of the expansion and 
the highest energy of the fit to be
\begin{eqnarray}
\alpha=1.15, \quad s_0=(1.1 \, {\rm GeV})^2.
\end{eqnarray}

For $s\geq{(m_{\eta}+m_K)^2}$, we will use once more the inelastic formalism of Eqs.\eqref{eq:inelunit},\eqref{eq:inelres}. Thus, we  write
\begin{equation}
t_1^{1/2}(s)=\frac{S^r_1 S^r_2 S^r_3-1}{2i\sigma(s)},
\end{equation}
where all the $S_k^r$ are of the form in Eq.\eqref{eq:inelres}, with
\begin{eqnarray}
P_1&=&(s_{r1}-s)\beta+e_1G_1 \frac{p_1(q_{\pi K})}{p_1(q_{\pi K}^r)}\frac{q_{\pi K}^2-\hat{q}_{\pi K}^2}{(q^r_{\pi K})^2-\hat{q}_{\pi K}^2}\frac{q_{\pi K}}{q^r_{\pi K}},\nonumber\\
P_{2,3}&=&e_{2,3}G_{2,3} \frac{p_{2,3}(q_{\pi K})}{p_{2,3}(q_{\pi K}^r)}\frac{q_{\pi K}^2-\hat{q}_{\pi K}^2}{(q^r_{\pi K})^2-\hat{q}_{\pi K}^2}\frac{q_{\pi K}}{q^r_{\pi K}},\\
Q_{1,2,3}&=&(1-e_{1,2,3})G_{1,2,3} \frac{p_{1,2,3}(q_{\pi K})}{p_{1,2,3}(q_{\pi K}^r)}\left(\frac{q_{\eta K}}{q^r_{\eta K}}\right)^3\Theta_{\eta K}(s).
\nonumber
\end{eqnarray}
In addition, 
\begin{equation}
p_i (q_{\pi K})=1+a_i q_{\pi K}^2,
\end{equation}
and $\Theta_{\eta K}(s)=\Theta(s-(m_K+m_\eta)^2)$ is the step function at the $K\eta$ threshold.
Again, in order to impose continuity at $K\eta$ threshold
we have defined
$\beta\equiv 1/\cot\delta_1^{1/2}((m_K+m_\eta)^2)$,
with $\delta_1^{1/2}$ calculated from the elastic expression in Eq.\eqref{eq:cot121}. 

\subsubsection{$t_P$-wave}

\begin{figure}
\centering
\includegraphics[scale=0.33]{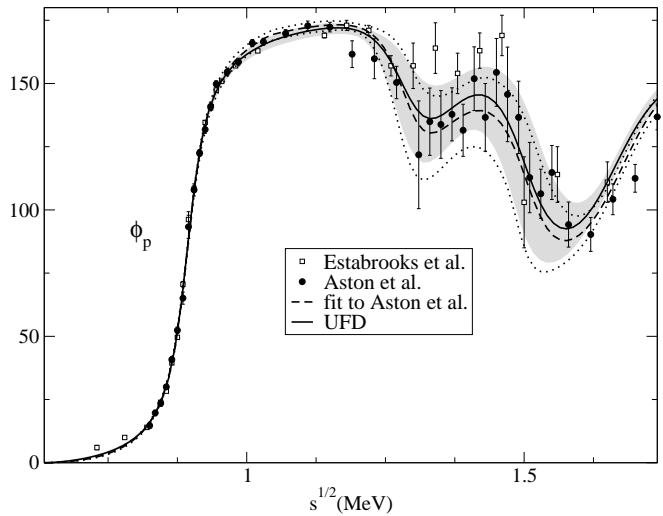}
 \caption{\rm \label{fig:pwavephaseufd} 
Data on $\phi_P(s)$ from 
Estabrooks et al. \cite{Estabrooks:1977xe} and Aston et al.
 \cite{Aston:1987ir}. The continuous line is our unconstrained fit (UFD), 
whose uncertainties are covered by the gray band. 
For comparison we show as a dashed line
a fit only to the data from \cite{Aston:1987ir}, 
whose corresponding uncertainties
are delimited by the dotted lines.}
\end{figure}

Thus, now that we have the 
functional forms for the $I=1/2$ and $I=3/2$ $P$-waves, 
we can perform the fit to all the $P$-wave data. As we did for the $S$-wave we first define
\begin{equation}
t_P(s)=\vert t_P(s)\vert e^{i\phi_P(s)},
\end{equation}
which is sometimes used with the alternative normalization
\begin{equation}
\hat t_P(s)=t_P(s) \sigma(s).
\end{equation}
As commented before, we fit simultaneously the $I=3/2$ data in Fig.\ref{fig:p32wavephase}
and the data on both $\vert \hat t_P\vert$ and $\phi_P$ that 
we show in Figs.~\ref{fig:pwavemodufd} and \ref{fig:pwavephaseufd}.
Note that once again there are clear systematic deviations of certain points, particularly from the Estabrooks et al.
data set \cite{Estabrooks:1977xe}. In this case we have 
proceeded as follows: we have performed a first fit,  then we have added a systematic uncertainty
to the isolated incompatible data points, which is half of their distance to the central value of the fit. In regions where the two sets of data are incompatible a systematic uncertainty is also added to each set, which corresponds to half of the average difference from the fit to the data set in that region. With these additional systematic uncertainties 
we have performed a final fit, which we call Unconstrained Fit to Data (UFD),
with $\chi^2/d.o.f.=76.4/(78-12+1)$.
The resulting curves are also shown in Figs.\ref{fig:pwavemodufd} and \ref{fig:pwavephaseufd},
together with a fit in which we have only fitted the Aston et al. data for the $I=1/2$ wave. It can be noticed that in such case the result would still be compatible with our UFD.

\begin{figure}
\centering
\includegraphics[scale=0.33]{ModulusP.eps}
 \caption{\rm \label{fig:pwavemodufd} 
Data on $\vert \hat t_P(s)\vert$ from \cite{Estabrooks:1977xe}, \cite{Aston:1987ir}.
The continuous line is our unconstrained fit (UFD), 
whose uncertainties are covered by the gray band. 
For comparison we show as a dashed line
a fit only to the data from \cite{Aston:1987ir}, 
whose corresponding uncertainties
are delimited by the dotted lines.
}
\end{figure}

Once all $P$-wave data have been fitted, we can separate the different isospin components.
The $I=3/2$ UFD result was already discussed in a Subsec.\ref{subsec:P32UFD} and its
parameters were given in Table~\ref{tab:P32param}.

\begin{figure}
\centering
\includegraphics[scale=0.33]{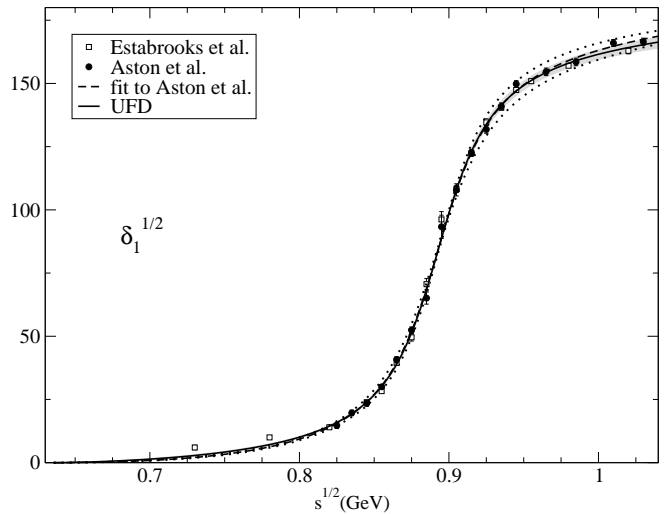}
 \caption{\rm \label{fig:swavephase} 
$P^{1/2}$-wave
phase shift below $K\eta$ threshold. 
The continuous line is our UFD parameterization whose uncertainty is covered by the gray band. 
Data from 
Estabrooks et al. \cite{Estabrooks:1977xe} and Aston et al.
 \cite{Aston:1987ir}.
The dashed curve is the fit to Aston et al. data alone
and the dotted lines cover its corresponding uncertainty.}
\end{figure}

Concerning the $P^{1/2}$-wave, let us first look at the elastic region.
When restricted below the $K\eta$ threshold
the UFD result has $\chi^2/d.o.f.=27/(34-4+1)$ and the corresponding parameters are listed in Table~\ref{tab:Pelparam}. The resulting curve for the $P^{1/2}$-wave can be seen in Fig.\ref{fig:swavephase}, where the distinct shape of the $K^*(892)$ is nicely observed. We are also showing a fit where only the data of Aston et al. has been fitted and how the results are hard to distinguish from our UFD line, except for the somewhat larger uncertainty band of the latter, particularly at higher energies.

\begin{table}[h] 
\caption{P$^{1/2}$-wave parameters in the elastic region.} 
\centering 
\begin{tabular}{c c c } 
\hline\hline  
\rule[-0.05cm]{0cm}{.35cm}Parameter & UFD & CFD \\ 
\hline 
\rule[-0.05cm]{0cm}{.35cm}$B_0$ & 0.97   $\pm$0.02        & 0.97   $\pm$0.02\\
\rule[-0.05cm]{0cm}{.35cm}$B_1$ & 0.98   $\pm$0.30        & 0.55   $\pm$0.30\\ 
\rule[-0.05cm]{0cm}{.35cm}$B_2$ & 0.79   $\pm$0.95        & 0.75   $\pm$0.95\\ 
\rule[-0.05cm]{0cm}{.35cm}$m_r$ & 0.8957 $\pm$0.0004GeV & 0.8957 $\pm$0.0004GeV\\
\hline 
\end{tabular} 
\label{tab:Pelparam} 
\end{table} 

The UFD parameters for the $P^{1/2}$-wave inelastic parameterization
are given in Table~\ref{tab:Pinpa}. 
Note that to describe the inelastic region we still 
need to take into account the high energy tail of 
the $K^*(892)$ resonance, which is elastic, so that we set $e_1=1$.
In addition its mass is fixed to the neutral case, 896 MeV, since this 
is the one measured in the LASS \cite{Aston:1987ir} and Estabrooks et al.  \cite{Estabrooks:1977xe}
experiments.
The other resonance shapes of the $K^*(1410)$ and $K^*(1680)$ are also very nicely described.

\begin{table}[h] 
\caption{P$^{1/2}$-wave parameters in the inelastic region.} 
\centering 
\begin{tabular}{c c c } 
\hline\hline  
\rule[-0.05cm]{0cm}{.35cm}Parameters & UFD & CFD \\ 
\hline 
\rule[-0.05cm]{0cm}{.35cm}$a_1$ & -1.90             $\pm$0.10GeV$^{-2}$  & -1.76             $\pm$0.10GeV$^{-2}$  \\ 
\rule[-0.05cm]{0cm}{.35cm}$a_2$ & -2.14             $\pm$0.23GeV$^{-2}$   &  -2.33             $\pm$0.23GeV$^{-2}$  \\ 
\rule[-0.05cm]{0cm}{.35cm}$a_3$ & -1.34             $\pm$0.07GeV$^{-4}$   &  -1.41            $\pm$0.07GeV$^{-4}$\\ 
\rule[-0.05cm]{0cm}{.35cm}$\sqrt{s_{r1}}$ & 0.896  GeV (fixed)             &  0.896  GeV (fixed)    \\ 
\rule[-0.05cm]{0cm}{.35cm}$\sqrt{s_{r2}}$ & 1.346 $\pm$0.012GeV     &  1.347 $\pm$0.012GeV \\ 
\rule[-0.05cm]{0cm}{.35cm}$\sqrt{s_{r3}}$ & 1.644 $\pm$0.005GeV     &  1.645 $\pm$0.005GeV \\ 
\rule[-0.05cm]{0cm}{.35cm}$e_1$ & 1   (fixed)                                & 1 (fixed) \\ 
\rule[-0.05cm]{0cm}{.35cm}$e_2$ & 0.052           $\pm$0.007          & 0.055           $\pm$0.007 \\ 
\rule[-0.05cm]{0cm}{.35cm}$e_3$ & 0.295           $\pm$0.016          & 0.306           $\pm$0.016\\ 
\rule[-0.05cm]{0cm}{.35cm}$G_1$ & 0.044           $\pm$0.003GeV     & 0.044           $\pm$0.003GeV  \\ 
\rule[-0.05cm]{0cm}{.35cm}$G_2$ & 0.217           $\pm$0.041GeV     & 0.231           $\pm$0.041GeV \\ 
\rule[-0.05cm]{0cm}{.35cm}$G_3$ & 0.295           $\pm$0.018GeV     & 0.306           $\pm$0.018GeV  \\ 
\hline 
\end{tabular} 
\label{tab:Pinpa} 
\end{table} 

Let us remark that there is a recent fit to the $t_P$ data \cite{Guo:2011aa}, neglecting the $I=3/2$ wave as usual, in which
the authors also consider three poles for the $I=1/2$ partial wave
within a two-channel K-matrix approach, 
the channels being $\pi K$ and $\pi K^*(892)$. 
In \cite{Guo:2011aa} only the central value of the fit is given and,
since it is a fit to basically the same data we fit here, the results are relatively similar to ours within uncertainties, actually, around 1 GeV it is slightly closer to our CFD result, that we will discuss later on, than to the UFD result discussed here. 
Note also that the parameterization in \cite{Guo:2011aa} is a fit to data up to 1.8 GeV and that, in principle, it could be extrapolated up to 2.3 GeV.

\subsection {$D$-waves}
\subsubsection {I=3/2 $D$-wave}

\begin{figure}
\centering
\includegraphics[scale=0.33]{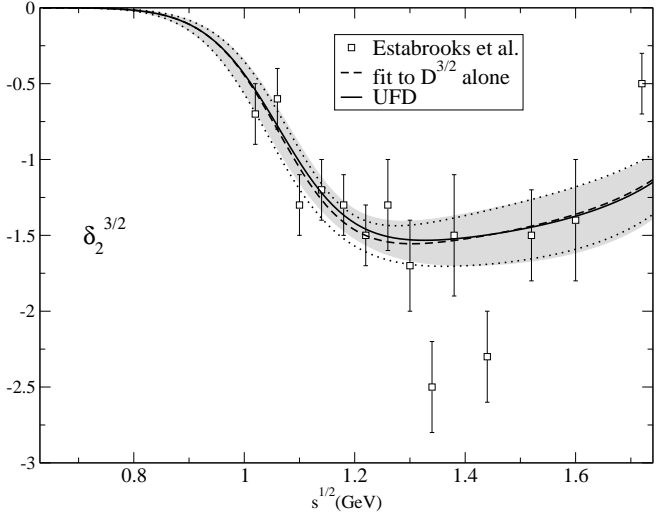}
 \caption{\rm \label{fig:D32phase} 
Data on the $I=3/2$ $D$-wave from Estabrooks et al. \cite{Estabrooks:1977xe}.
We also show our UFD result as a solid line with a gray uncertainty band, 
which is obtained by fitting this data in a simultaneous fit with the data on the $t^{1/2}_2+t^{3/2}_2/2$ combination.
For comparison we show with a dashed line a fit only to the data in this figure.
}
\end{figure}

Once again, only Estabrooks et al. \cite{Estabrooks:1977xe} provide data for the $I=3/2$ $D$-wave phase shift up to 1.74 GeV, 
which we show in Fig.\ref{fig:D32phase}. Note it is very small in the whole energy region.
No inelasticity has been measured so that we can use the elastic formalism parameterized with the conformal expansion
 in Eqs.\eqref{eq:generalconformal} and \eqref{eq:conformalvars}, as follows:
\begin{equation}
\cot\delta_1^{3/2}(s)=\frac{\sqrt{s}}{2q^5}(B_0+B_1\omega+B_2\omega^2).
\end{equation}
Three conformal parameters are enough to describe this wave.
As we did for the $P^{3/2}$-wave, the $\alpha$ parameter
that defines the conformal variable $\omega$ in  Eq.\eqref{eq:conformalvars} has been chosen
so that the center of the conformal expansion lies at the center of the region where data exists. 
Thus, for this wave we have set
\begin{eqnarray}
\alpha=1.45, \quad s_0=(1.84 \, {\rm GeV})^2.
\end{eqnarray}

As it can be noticed in Fig.\ref{fig:D32phase}, there are sizable systematic uncertainties, 
which can be simply taken into account by multiplying the statistical uncertainties by $\sqrt{2}$.
The resulting fit yields a $\chi^2/d.o.f\simeq1.1$. However, as it happened with the $S^{3/2}$ and $P^{3/2}$-waves, 
our final fit for the $I={3/2}$ D-wave is obtained from a simultaneous fit
to the data for this wave alone together with the
data on the $t_D \equiv t^{1/2}_2+t^{3/2}_2/2$ combination obtained
by Estabrooks et al. \cite{Estabrooks:1977xe} and Aston et al. \cite{Aston:1987ir}.
The parameters of such 
Unconstrained Fit to Data (UFD) are given in Table.\ref{tab:D32param} 
and we show the resulting phase shift as a continuous line in Fig.\ref{fig:D32phase},
where the gray band covers the corresponding uncertainty. 

\begin{table}[h] 
\caption{Parameters of the D$^{3/2}$-wave.} 
\centering 
\begin{tabular}{c c c } 
\hline\hline  
\rule[-0.05cm]{0cm}{.35cm}Parameter & UFD & CFD \\ 
\hline 
\rule[-0.05cm]{0cm}{.35cm}$B_0$ & -1.70  $\pm$0.12  & -1.67  $\pm$0.12\\
\rule[-0.05cm]{0cm}{.35cm}$B_1$ & -6.5  $\pm$1.7  & -7.0  $\pm$1.7\\ 
\rule[-0.05cm]{0cm}{.35cm}$B_2$ & -36 $\pm$9  & -38$\pm$9\\ 
\hline 
\end{tabular} 
\label{tab:D32param} 
\end{table}

In Fig.\ref{fig:D32phase}  we also show, as a dashed line, the result when fitting only the data in that figure and not data on the $t_D$ combination.
The corresponding uncertainty band is delimited by dotted lines. As we can see it is 
very similar to our UFD curve.

\subsubsection {I=1/2 $D$-wave}

As it happened with the $S$ and $P$-waves, 
the $I$=1/2 $D$-wave is only measured together
with the $I$=3/2-wave in the 
$t_D \equiv t^{1/2}_2+t^{3/2}_2/2$
combination. 
In the literature it is usual to 
neglect the $D^{3/2}$-wave, because as we have just seen is very small, 
but we will keep it in our fits for completeness.

Let us then describe our fit to the $D^{1/2}$-wave, which
is dominated by the  $K_2^*(1430)$ resonance, whose branching ratio to $\pi K$ is 
approximately 50\%, so that we have to use an inelastic formalism
as in Eqs.\eqref{eq:inelunit}, \eqref{eq:inelback}, \eqref{eq:inelres}.
In practice, it is enough to consider a non-resonant background and a resonant-like form,
as follows:
\begin{equation}
t_2^{1/2}=\frac{S^b_0S^r_1-1}{2i\sigma(s)},
\label{eq:Dwave}
\end{equation}
with a background term
\begin{equation}
S^b_0=e^{2i(p(s))},
\end{equation}
where 
\begin{eqnarray}
p(s)&=&\phi_0q_{\eta K}^5\Theta_{\eta K}(s)
+q_{\eta' K}^5(\phi_1+\phi_2q_{\eta' K}^2)\Theta_{\eta' K}(s),\nonumber
\end{eqnarray}
and $\Theta_{ab}=\Theta(s-(m_a+m_b)^2)$.
A resonant term is also considered in order to describe 
easily the $K_2^*(1430)$ shape, namely
\begin{eqnarray}
S^r_1&=&\frac{s_{r1}-s+i(P_1-Q_1)}{s_{r1}-s-i(P_1+Q_1)},\\
P_1&=&e_1G_1 \frac{p_1(q_{\pi K})}{p_1(q_{\pi K}^r)}\left(\frac{q_{\pi K}}{q_{\pi K, r}}\right)^5,\nonumber\\
Q_1&=&(1-e_1)G_1 \frac{p_1(q_{\pi K})}{p_1(q_{\pi K}^r)}\left(\frac{q_{\eta K}}{q_{\eta K, r}}\right)^5\Theta_{\eta K}(s),\nonumber
\end{eqnarray}
with $p_1(q_{\pi K})=1+a q_{\pi K}^2$.

\subsubsection{$t_D$-wave}

\begin{figure}
\centering
\includegraphics[scale=0.33]{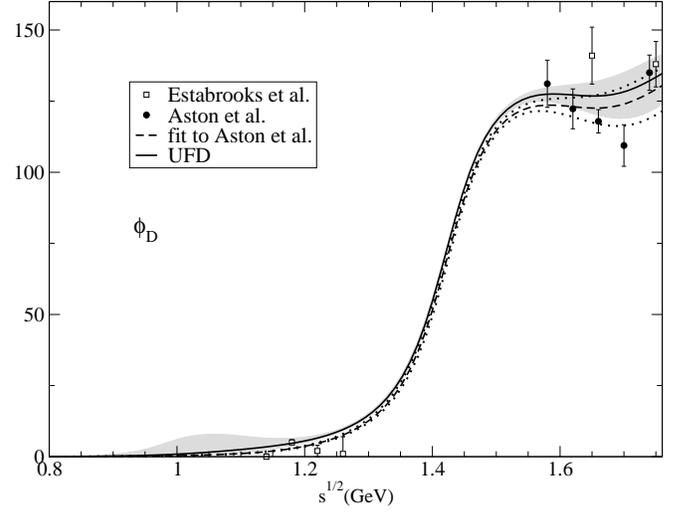}
 \caption{\rm \label{fig:dwavephase} 
Data on $\phi_D(s)$ from \cite{Estabrooks:1977xe,Aston:1987ir}. The continuous line is our unconstrained fit (UFD), 
whose uncertainties are covered by the gray band. 
For comparison we show, as a dashed line, a fit only to the data from \cite{Aston:1987ir} whose corresponding uncertainties
are delimited by the dotted lines. }
\end{figure}

Once more we define
\begin{equation}
t_D(s)\equiv \vert t_D(s)\vert e^{i\phi_D(s)},
\quad \hat t_D(s)=t_D(s) \sigma(s). 
\end{equation}
Thus, 
in Figs.~\ref{fig:dwavemod} and \ref{fig:dwavephase}  we show
the data on $\vert \hat t_D \vert$ and $\Phi_D$, respectively.
As we did for the $P$-wave, we have treated the systematic uncertainties 
as follows: we have performed a first fit and added a systematic uncertainty
to those isolated data points the are incompatible with it. 
This systematic uncertainty is half of their distance to the central value of the fit. 
In regions where the two sets of data are incompatible the systematic uncertainty
is half of the average difference from the fit to each set in that region. With these additional systematic uncertainties 
we have performed a final fit, called Unconstrained Fit to Data (UFD), which is shown 
as a continuous line in Figs.\ref{fig:dwavemod} and \ref{fig:dwavephase}.
The UFD uncertainty is represented as a gray band. The total $\chi^2/d.o.f.$ is $49/(44-6+1)$.
In addition, we show as a dashed line the result that 
is obtained if only the data on $\hat t_D$ from Aston et al.\cite{Aston:1987ir} is fitted. The central values are almost indistinguishable but the uncertainties are smaller. We still
prefer our UFD parameterization because it contains more experimental information, although the uncertainties come larger due to the systematic uncertainties that we have taken into account in our UFD set. 
The UFD parameters are shown in Table~\ref{tab:Dwave}.

\begin{table}[h] 
\caption{Parameters of the D$^{1/2}$ fit.} 
\centering 
\begin{tabular}{c c c } 
\hline\hline  
\rule[-0.05cm]{0cm}{.35cm}Parameters & UFD & CFD \\ 
\hline 
\rule[-0.05cm]{0cm}{.35cm}$\phi_0$ & 2.17         $\pm$0.26 GeV$^{-5}$     &  3.00         $\pm$0.26 GeV$^{-5}$\\ 
\rule[-0.05cm]{0cm}{.35cm}$\phi_1$ & -12.1       $\pm$1.7GeV$^{-5}$      & -9.3       $\pm$1.7GeV$^{-5}$\\ 
\rule[-0.05cm]{0cm}{.35cm}$\sqrt{s_{r1}}$ & 1.446 $\pm$0.002GeV        &  1.445  $\pm$0.002GeV\\ 
\rule[-0.05cm]{0cm}{.35cm}$e_1$ & 0.466           $\pm$0.006             & 0.465           $\pm$0.006\\ 
\rule[-0.05cm]{0cm}{.35cm}$G_1$ & 0.220           $\pm$0.009GeV        & 0.222           $\pm$0.009GeV\\ 
\rule[-0.05cm]{0cm}{.35cm}$a$ & -0.53             $\pm$0.16GeV$^{-2}$      &-0.72             $\pm$0.16GeV$^{-2}$\\
\hline 
\end{tabular} 
\label{tab:Dwave} 
\end{table}

\begin{figure}
\centering
\includegraphics[scale=0.33]{Dmod.eps}
 \caption{\rm \label{fig:dwavemod} 
Data on $\vert \hat  t_D(s)\vert$ from \cite{Estabrooks:1977xe,Aston:1987ir}.
The continuous line is our unconstrained fit (UFD), 
whose uncertainties are covered by the gray band. 
For comparison we show, as a dashed line, a fit only to the data from  \cite{Aston:1987ir} whose corresponding uncertainties
are delimited by the dotted lines.
}
\end{figure}

\subsection {$F^{1/2}$-wave}

\begin{figure}
\centering
\includegraphics[scale=0.33]{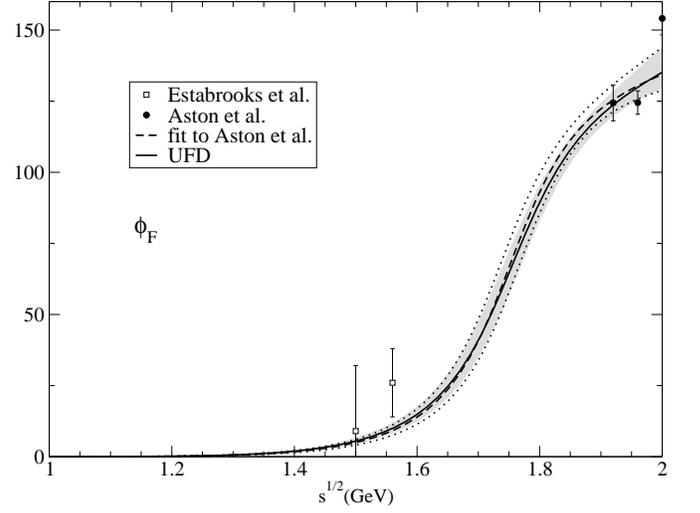}
 \caption{\rm \label{fig:fwavephase} 
Data on $\phi_F(s)$ from \cite{Estabrooks:1977xe,Aston:1987ir}. The continuous line is our unconstrained fit (UFD), 
whose uncertainties are covered by the gray band. 
For comparison we show, as a dashed line, a fit to the data from \cite{Aston:1987ir}
whose corresponding uncertainties
are delimited by the dotted lines.}
\end{figure}

 Once more we define 
\begin{equation}
t_F(s)\equiv \vert t_F(s)\vert e^{i\phi_F(s)},
\quad \hat t_F(s)=t_F(s) \sigma(s). 
\end{equation}
For this wave there are 
no observations of an $I=3/2$ channel, which is neglected in
the literature as will be done here too. 
In addition, the threshold suppression is so large that there are no data below 1.5 GeV as can be seen in Figs.\ref{fig:fwavemod} and \ref{fig:fwavephase}.
In the latter we can observe that there are only two data points with very large uncertainties for the phase $\phi_F$ below 1.85 GeV. Thus, in order to stabilize our fit we will extend
our data sample to 2 GeV, although later on we will only make use of our 
partial wave parameterizations up to 1.74 GeV.

The most salient feature of this wave is the 
$K_3^*(1780)$ resonance, whose branching ratio to $\pi K$ is slightly less than 20\%. Therefore we will
need the usual inelastic formalism explained in the introduction to this section:
\begin{equation}
t_3^{1/2}=\frac{S^r_1-1}{2i\sigma(s)},
\label{eq:Fwave}
\end{equation}
with
\begin{eqnarray}
S^r_1&=&\frac{s_{r1}-s+i(P_1-Q_1)}{s_{r1}-s-i(P_1+Q_1)},\\
P_1&=&e_1G_1 \frac{p_1(q_{\pi K})}{p_1(q_{\pi K}^r)}\left(\frac{q_{\pi k}}{q_{\pi k, r}}\right)^{\!\!7},\nonumber\\
Q_1&=&(1-e_1)G_1 \frac{p_1(q_{\pi K})}{p_1(q_{\pi K}^r)}\left(\frac{q_{\eta k}}{q_{\eta k, r}}\right)^{\!\!7} \Theta_{\eta K}(s).
\nonumber \end{eqnarray}
In addition, $p_1(q_{\pi K})=1+a q_{\pi K}^2$ and $\Theta_{\eta K}(s)=\Theta(s-(m_\eta+m_K)^2)$.

No background term is needed for this wave because 
its behavior is well described using the resonant-like form only,
 as it can be observed in Fig.\ref{fig:fwavemod} and Fig.\ref{fig:fwavephase}.
The fit yields $\chi^2/d.o.f.=16/(21-4+1)$ 
and the UFD parameters listed in Table~\ref{tab:Fwave}.

\begin{table}[h] 
\caption{Parameters of the $F^{1/2}$-wave.} 
\centering 
\begin{tabular}{c c c } 
\hline\hline  
\rule[-0.05cm]{0cm}{.35cm}Parameters & UFD & CFD \\ 
\hline 
\rule[-0.05cm]{0cm}{.35cm}$\sqrt{s_{r1}}$ & 1.801 $\pm$0.013GeV    &   1.804          $\pm$0.013GeV\\ 
\rule[-0.05cm]{0cm}{.35cm}$e_1$ & 0.181           $\pm$0.006         &  0.184           $\pm$0.006\\ 
\rule[-0.05cm]{0cm}{.35cm}$G_1$ & 0.47           $\pm$0.05GeV    &  0.50           $\pm$0.05GeV\\ 
\rule[-0.05cm]{0cm}{.35cm}$a$ & -0.88            $\pm$0.10GeV$^{-2}$ & -0.97           $\pm$0.10GeV$^{-2}$\\
\hline 
\end{tabular} 
\label{tab:Fwave} 
\end{table}

\begin{figure}
\centering
\includegraphics[scale=0.33]{Fmod.eps}
 \caption{\rm \label{fig:fwavemod} 
Data on $\vert \hat t_F(s)\vert$ from \cite{Estabrooks:1977xe,Aston:1987ir}.
The continuous line is our unconstrained fit (UFD), 
whose uncertainties are covered by the gray band. 
For comparison we show, as a dashed line, a fit to the data from \cite{Aston:1987ir}
whose corresponding uncertainties
are delimited by the dotted lines.
}
\end{figure}

\subsection{Regge Parameterization}
\label{sec:Regge}

There are no data on $I=3/2$ above 1.74 GeV, thus above that energy
we will make use of the Regge parameterization for $\pi K$ scattering
in \cite{Pelaez:2004vs,GarciaMartin:2011cn}, which was obtained from factorization after fitting
data on $NN$, $N\pi$, $NK$ and $\pi\pi$ high energy scattering.
Note that 
for $\pi K$ scattering only 
the exchange of isospin 0 and 1 can occur in the $t$ channel.

For the isoscalar exchange there are two contributions: the Pomeron, called $P(s)$ here,
and the subleading $f_2$ trajectory, called $P'(s)$, so that we write:
\begin{equation}
\im T^{(I_t=0)}_{\pi K}(s,t)=f_{ K/\pi}\left[P(s,t)+rP'(s,t)\right],
\end{equation}
where
\begin{eqnarray}
P(s,t)&=&\beta_P\psi_P(t)\alpha_P(t)\frac{1+\alpha_P(t)}{2}e^{bt}\left(\frac{s}{s'}\right)^{\alpha_P(t)},\nonumber\\
P'(s,t)&=&\beta_{P'}\psi_{P'}(t)\frac{\alpha_{P'}(t)(1+\alpha_P(t))}{\alpha_{P'}(0)(1+\alpha_P(0))}e^{bt}\left(\frac{s}{s'}\right)^{\alpha_{P'}(t)},\nonumber\\
\alpha_P(t)&=&1+t\alpha'_P, \psi_P=1+c_Pt,\nonumber\\
\alpha_{P'}(t)&=&\alpha_{P'}(0)+t\alpha'_{P'}, \psi_P=1+c_{P'}t.
\end{eqnarray}
Note that, by using factorization, the substitution of  the
$\pi\pi$-Pomeron vertex by the $KK$-Pomeron vertex is taken into account
by the $f_{K/\pi}$ constant, whereas $r f_{K/\pi}$ takes care of the similar factorization for $P'$.

Since in this work we are interested in Forward Dispersion Relations, we will only use the above Regge formulas with $t=0$, but we provide the full expressions for completeness.

For the isovector exchange only the $\rho$ trajectory is needed and we use
\begin{equation}
\im T^{(I_t=1)}_{\pi K}(s,t)=g_{ K/\pi}\im T^{(I_t=1)}_{\pi\pi}(s,t),
\end{equation}
with 
\begin{eqnarray}
\im T^{(I_t=1)}_{\pi \pi}(s,t)&=&\beta_{\rho}\frac{1+\alpha_{\rho}(t)}{1+\alpha_{\rho}(0)}\varphi(t)e^{bt}\left(\frac{s}{s'}\right)^{\alpha_{\rho}(t)},\nonumber\\
\alpha_{\rho}(t)& =&\alpha_{\rho}(0)+t\alpha'_{\rho}+\frac{1}{2}t^2\alpha''_{\rho},\nonumber\\
\varphi(t)& =&1+d_{\rho}t+e_{\rho}t^2.
\end{eqnarray}
Once again, the replacement of the $\pi\pi\rho$ vertex by the $KK\rho$ one 
is described by the $g_{ K/\pi}$ constant, assuming factorization.
Note that in \cite{Pelaez:2004vs} the value of $g_{ K/\pi}=1.1\pm0.1$ 
was used together with $\beta_{\rho}=0.94\pm0.20$ to provide a description of $\pi K$.
However, the same group \cite{GarciaMartin:2011cn} updated
their $\pi\pi$ scattering analysis using dispersion relations and $\pi\pi$ scattering data at high energies to find 
$\beta_{\rho}=1.48\pm0.14$. Consequently, if we want to use this latter value, we also have to update $g_{ K/\pi}=0.70\pm0.07$. One should nevertheless take into account that the information on this parameter is relatively scarce, since, in contrast to $\pi\pi$ scattering, 
there are no high energy data on $\pi K$ scattering. Thus it has to be determined only from factorization of $K N$ scattering.
Note that $\beta_{\rho}$, which is the equivalent value for $\pi\pi$,
suffered a large revision when taking into account dispersion relations. 
Thus, large deviations in $g_{ K/\pi}$ should not come as a surprise
and they actually do occur when imposing our dispersive constraints on $\pi K$ scattering.

The set of Regge parameters used before imposing any $\pi K$ dispersion relation 
will be labeled  as ``Unconstrained Fit to Data''  (UFD) values, similarly to what we have been doing so far with our 
partial wave parameterizations. Correspondingly, we will refer to "Constrained Fit to Data" (CFD) values when in the next sections Forward Dispersion Relations will be imposed on our fits.
Those Regge parameters that could be determined without $K \pi$ input
are shown in Table~\ref{tab:regge} and their values are fixed both for the UFD and CFD parameterizations.
They just correspond to the values in the original  works \cite{Pelaez:2004vs,GarciaMartin:2011cn}.

\begin{table}[h]
\caption{Values of Regge parameters obtained in \cite{Pelaez:2004vs,GarciaMartin:2011cn}. Since these could be fixed using reactions other than $\pi K$ scattering,
they will be fixed both in our UFD and CFD parameterizations.
\label{tab:regge} } 
\centering 
\begin{tabular}{c c} 
\hline\hline  
\rule[-0.15cm]{0cm}{.55cm} Regge & Used both for \\ 
\rule[-0.15cm]{0cm}{.55cm} Parameters & UFD and CFD \\ 
\hline 
$s'$ & 1 GeV$^{2}$                                  \\
$b$ & 2.4                 $\pm$0.5 GeV$^{-2}$       \\ 
$\alpha'_{\rho}$ & 0.9 GeV$^{-2}$                   \\
$\alpha''_{\rho}$ & -0.3 GeV$^{-4}$                 \\ 
$d_{\rho}$ & 2.4          $\pm$0.5 GeV$^{-2}$       \\
$e_{\rho}$ & 2.7          $\pm$2.5                  \\
$\alpha'_{P}$ & 0.2       $\pm$0.1 GeV$^{-2}$       \\
$\alpha'_{P'}$ & 0.9 GeV$^{-2}$                     \\ 
$c_{P}$ & 0.6             $\pm$1 GeV$^{-2}$         \\
$c_{P'}$ & -0.38          $\pm$0.4 GeV$^{-2}$       \\ 
$\beta_{\rho}$ & 1.48     $\pm$0.14                 \\
$\alpha_{\rho}(0)$ & 0.53 $\pm$0.02                 \\ 
$\beta_{P}$ & 2.50        $\pm$0.04                 \\
$c_{P}(0)$ & 0            $\pm$0.04                 \\
$\beta_{P'}$ & 0.80       $\pm$0.05                 \\
$c_{P'}(0)$ & -0.4        $\pm$0.4                  \\
$\beta_{2}$ & 0.08        $\pm$0.2                  \\
$\alpha_{P'}(0)$ & 0.53   $\pm$0.02                 \\
\hline 
\end{tabular} 
\end{table} 

In contrast, the values 
$f_{ K/\pi}$, $g_{ K/\pi}$ and $r$, which are directly related to $\pi K$ scattering,
are listed in Table~\ref{tab:reggeKpi} and in this work they are allowed to vary from the UFD to the CFD parameterization, although in practice $r$ stays the same.

\begin{table}[h]
\caption{Values of Regge parameters directly related to $\pi K$ scattering.
In practice $r$ does not change from the UFD to CFD parameterization.
\label{tab:reggeKpi} } 
\centering 
\begin{tabular}{c c c} 
\hline\hline  
\rule[-0.15cm]{0cm}{.55cm} Parameters & UFD & CFD \\ 
\hline 
$f_{ K /\pi}$ & 0.67       $\pm$0.01                 & 0.66       $\pm$0.01\\
$r$ & 5$\cdot 10^{-2}$                              &  5$\cdot 10^{-2}$\\
$g_{ K/\pi}$ & 0.70        $\pm$0.07                &0.53        $\pm$0.07\\
\hline 
\end{tabular} 
\end{table}

\section{Forward Dispersion Relations and sum rules}

The aim of this work is to provide a simple set of partial 
waves which are consistent with basic requirements of analyticity (or causality) and crossing.
These features impose stringent constraints on the scattering amplitude,
which translate into integral equations 
that relate the $\pi K$ scattering amplitude at a given energy with 
an integral over the whole physical energy region. 
In this section we introduce a complete set of Forward Dispersion Relations
that will be used first to check the consistency of  our parameterizations
and next as constraints on our fits.

\subsection{Forward dispersion relations}

Forward Dispersion Relations (FDR), i.e., calculated at $t=0$, are useful because
forward scattering is relatively easy to measure in the whole energy region, 
since it is related to the total cross section by the optical theorem.
Moreover, this is the only fixed value of $t$ for which the integrands in the dispersion relation will be given directly in terms of the imaginary part of a physical amplitude.
They are applicable at any energy, in contrast to Roy-like equations which, in practice, 
have a limited applicability energy range
due to the projection in partial waves.

Fixed-$t$ dispersion relations for $\pi K$ have been 
frequently used in the literature as an intermediate step for
the derivation of more elaborated dispersion relations for partial waves \cite{Buettiker:2003pp,Ader:1974yk,Palou:1974ma, Palou:1975uu,Steiner:1971ms}, or of sum rules for low energy parameters \cite{Ananthanarayan:2001uy},
but not directly as constraints on the amplitudes, as will be done here.

For the sake of simplicity, given that $s+t+u=2(m_K^2+m_\pi^2)$ and $t=0$, 
it is customary to use an abbreviated notation $T(s,t=0,u)=T(s)$.
It is also very convenient to make use of $s\leftrightarrow u$ crossing
to change the amplitudes from the isospin basis to 
the $s\leftrightarrow u$ symmetric and antisymmetric amplitudes. 
These are defined, respectively,  as
\begin{eqnarray}
T^+(s)&=&\frac{T^{1/2}(s)+2T^{3/2}(s)}{3}=\frac{T^{I_t=0}(s)}{\sqrt{6}},\nonumber\\
T^-(s)&=&\frac{T^{1/2}(s)-T^{3/2}(s)}{3}=\frac{T^{I_t=1}(s)}{2}.
\end{eqnarray}
In the last step we have indicated that these $T^+$ and $T^{-}$
correspond, by crossing, to the exchange of isospin 0 or 1 in the $t$-channel, respectively. This is relevant because it means that $T^+$ is dominated at high energies
by the $t$-channel exchanges
of the Pomeron and $P'$ trajectories, with no $\rho$ trajectory contribution, 
whereas the opposite occurs for $T^-$.

Since dispersive integrals extend to infinity, naively one would need 
two subtractions to ensure the convergence of the Pomeron contribution and one for that of the $\rho$ trajectory. For this reason, even if only used as intermediate steps for the derivation of other dispersion relations, the fixed-$t$ dispersion relations for $T^+$ are customarily written with two subtractions and those for $T^-$ at least with one.
However, this is not necessary, because the  $T^{\pm}$ FDR have 
integrals over the right-hand and left-hand cuts whose leading terms multiplying the Regge trajectories cancel against each other due to the symmetry properties. 
As a consequence, one subtraction  is enough to ensure the convergence of the $T^+$ FDR
and no subtraction is needed for the $T^-$ FDR. This kind of cancellations 
have been recently used for $\pi\pi$ scattering FDR in \cite{GarciaMartin:2011cn,Pelaez:2004vs,Kaminski:2006qe,Kaminski:2006yv}. 
Having more subtractions implies that the dispersion relation is determined up to a polynomial of higher order.
Thus, generically, less subtractions 
are convenient to avoid the propagation of the uncertainties in the subtraction constants
to become too large in the resonance region, whereas more subtractions are useful when concentrating on the threshold region. Since in this work we will deal with scattering up to 1.74 GeV, we will make use of FDR with the minimum number of subtractions needed, 
which also makes the equations slightly simpler.

Thus, for $T^+$ the once-subtracted FDR reads:
\begin{widetext}
\begin{eqnarray}
\re T^+(s)=T^+(s_{th})
+\frac{(s-s_{th})}{\pi}P\!\int^{\infty}_{s_{th}}\!ds'\left[\frac{\im T^+(s')}{(s'-s)(s'-s_{th})}
-\frac{\im T^+(s')}{(s'+s-2\Sigma_{\pi K})(s'+s_{th}-2\Sigma_{\pi K})}\right],
\label{eq:FDRTsym}
\end{eqnarray}
\end{widetext}
where $s_{th}=(m_\pi+m_K)^2$ and $P$ stands for the principal part of the integral. In contrast, 
the unsubtracted FDR for $T^-$ reads:
\begin{equation}
\re T^-(s)=
\frac{(2s-2\Sigma_{\pi K})}{\pi}P\!\!\int^{\infty}_{s_{th}}\!\!\!\! ds'
\frac{\im T^-(s')}{(s'-s)(s'+s-2\Sigma_{\pi K})}.
\label{eq:FDRTan}
\end{equation}

We have evaluated these two FDRs
at $50$ 
values of $\sqrt{s_i}$
 equally spaced between a minimum energy $\sqrt{s_{min}}$
 and 1.74 GeV, using as input for the integrals our UFD partial waves at $s'\leq 1.74\,$ GeV 
and the Regge UFD parameterizations above.
At each of these $\sqrt{s_i}$ we have also calculated the difference $d_i$
between 
the left and right hand sides of each FDR as well as its corresponding uncertainty $\Delta d_i$.
When $d_i\lesssim\Delta d_i$ we can consider that the FDR is satisfied within uncertainties
at the energy $\sqrt{s_i}$.

As a word of caution, let us remark that the uncertainties $\Delta d_i$ are calculated 
as the quadratic addition of the uncertainties due to the error bar of each parameter in the 
UFD parameterization. Note however that in the full physical amplitude, being a solution of the FDRs,
all these parameters would be correlated. Our parameterizations are just a fit to partial waves, 
many of which have been measured independently from one another and these correlations may be lost.
Therefore our FDR error bands only reflect the propagation of the uncertainties from our data parameterizations, without the possible correlations between parameters that may exist.

\begin{figure}
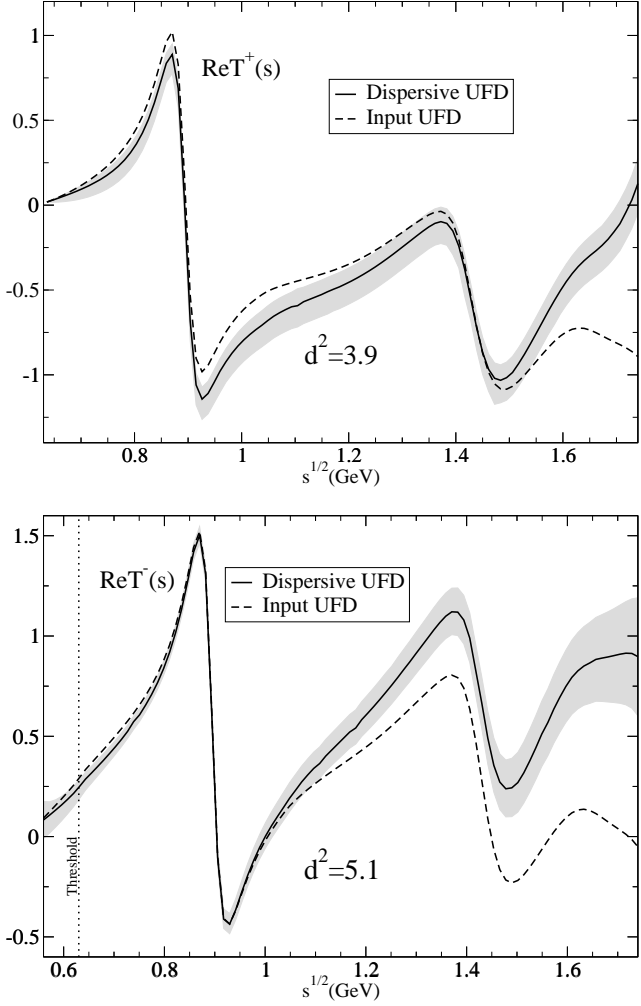

\centering
\includegraphics[scale=0.32]{Unconstrainedpar.eps}\vspace*{.3cm}
\includegraphics[scale=0.32]{Unconstrainedimpar.eps}
\caption{\rm \label{fig:FDRUFD} 
Comparison between the input (dashed) 
and the output for the $T^+$ (top) and $T^-$(bottom) FDRs when using
the UFD set. 
These correspond to the left hand side versus the right hand side
in Eqs.\eqref{eq:FDRTsym} and \eqref{eq:FDRTan}.
The gray bands describe the uncertainty of the difference.
}
\end{figure}

The results of the above calculation for  $T^+$ are shown in the upper panel of Fig.\ref{fig:FDRUFD}.
We plot Re$\,T^+$ calculated directly from the UFD parameterization (Input UFD)
versus Re$\,T^+$ calculated from the dispersive representation in Eq.\eqref{eq:FDRTsym} above (Dispersive UFD).  The gray area corresponds to 
adding $\pm\Delta d_i$ to the "Dispersive UFD" curve.
Note that for this symmetric amplitude we have set $\sqrt{s_{min}}$ at $K\pi$ threshold.
We can see that the Input UFD lies slightly beyond the uncertainty band up to 1.2 GeV,
but that it is  
much more separated beyond 1.55 GeV. The consistency of the data with the $T^+$ FDR is therefore not very satisfactory, particularly at higher energies.

The results for $T^-$ are plotted in the lower panel of Fig.\ref{fig:FDRUFD},
using the same conventions. In this case  we have set $\sqrt{s_{min}}=0.56\,$GeV, below threshold,
because the $T^-$ FDR has no subtractions and thus provides strong constraints
on a combination of scalar scattering lengths. The figure shows that the separation between both
calculations is slightly above $\Delta d_i$ up to 0.8 GeV. Above this energy, the $T^-$ FDR is nicely satisfied within uncertainties up to 1.2 GeV, where the difference between
the two curves starts growing, becoming much larger than $\Delta d_i$.  As we will see, the deviation at energies above 1.2 GeV is mainly caused by 
the  $\rho$-exchange Regge contribution.

In order to provide a quantitative measure of the fulfillment of each FDRs,
we have defined an averaged squared distance for each FDR,
\begin{equation} 
d_{T\pm}^2=\frac{1}{N}\sum_{i=1}^{N} \left(\frac{d_i}{\Delta d_i}\right)_{T^\pm}^2,
\label{eq:distances}
\end{equation}
which is rather similar to the usual definition of a $\chi^2$ function.
Consistency of the data parameterizations with FDRs would demand $d_{T\pm}^2\lesssim 1$.

In Table~\ref{tab:d2FDR} we show the values of $d_{T\pm}^2$ 
in different energy regions.
For the UFD set it is clear that the consistency with FDRs is not very satisfactory, particularly
in the inelastic region, and very inconsistent above 1.6 GeV. 
There is room for a considerable improvement that will be achieved in Sec.\ref{sec:CFD}, to the point of obtaining a constrained set of parameterizations (CFD) remarkably consistent with both FDRs up to 1.6. However, we will see that above that energy we are only able to improve the agreement but not achieve 
consistency within uncertainties.

\begin{table} 
\caption{Fulfillment of Forward Dispersion Relations (FDR) in different energy regions. We provide the averaged square distance
divided by relative error between the left and right hand sides of each FDR. 
Note the remarkable improvement from the UFD to the CFD parameterizations.} 
\centering 
\begin{tabular}{c | c c | c c} 
\hline\hline  
\rule[-0.15cm]{0cm}{.55cm} & \multicolumn{2}{c|}{UFD} & \multicolumn{2}{c}{CFD}\\ 
\rule[-0.15cm]{0cm}{.55cm}& $d_{T^+}^2$ & $d_{T^-}^2$& $d_{T^+}^2$ & $d_{T^-}^2$\\
\hline 
\rule[-0.15cm]{0cm}{.55cm} $\sqrt{s_{min}}\leq\sqrt{s}\leq m_K+m_\eta$&3.35  & 0.97& 0.39& 0.13\\
\rule[-0.15cm]{0cm}{.55cm} $m_K+m_\eta \leq\sqrt{s}\leq 1.6\,$GeV& 1.3& 6.8& 0.17& 0.70\\
\rule[-0.15cm]{0cm}{.55cm} $1.6\,$GeV$\leq\sqrt{s}\leq 1.74\,$GeV& 14.6& 12.8& 8.0& 0.5\\
\hline 
\rule[-0.15cm]{0cm}{.55cm} $\sqrt{s_{min}}\leq\sqrt{s}\leq 1.74\,$GeV& 3.9& 5.1& 1.3& 0.44\\
\hline 
\end{tabular} 
\label{tab:d2FDR} 
\end{table}

\subsection{Sum Rules for threshold parameters}

Threshold parameters of partial waves, 
defined in Eq.\eqref{eq:definitionab}, are of interest for our understanding
of the lowest energy Physics 
and particularly for studies within ChPT \cite{Gasser:1984gg}. 
In this section we present three sum rules (SR) that
provide a more
accurate determination of
certain combinations of threshold parameters, in terms of integrals,
than would be achieved directly from the partial wave parameterizations. 
These SR will be used first 
as tests of our UFD parameterizations 
and in Sec.\ref{sec:CFD} will be used as constraints.

Other sum rules have been derived for determining the ChPT low energy constants \cite{Ananthanarayan:2001uy}, but here we will use our own sum rules for threshold parameters.
In \cite{Buettiker:2003pp} a sum rule for a combination of scattering lengths is given, but it needs $\pi\pi\rightarrow K\bar K$ scattering input, and here we only want to use data on $\pi K$ scattering.

Thus, the first of our sum rules yields precisely the combination of scalar scattering lengths
measured at DIRAC \cite{Adeva:2014xtx}, see Eq.\eqref{eq:dirac} above. 
It is  basically the $T^-$ FDR evaluated at threshold
and for convenience we will write it as follows:
\begin{equation}
0=\Delta_a\equiv D_a-SR_a,\label{eq:SRa}
\end{equation}
where 
\begin{equation}
D_a\equiv\frac{1}{3}\left(a^{1/2}_0-a^{3/2}_0\right)
\end{equation}
is calculated ``directly'' from our parameterizations.
In principle it should be equal to the following integral expression:
\begin{equation}
SR_a\equiv\frac{2*m_\pi m_K}{\sqrt{s_{th}}} P\!\!\int^{\infty}_{s_{th}}{\frac{\im T^-(s')}{s'(s'-s_{th})}ds'}.\nonumber
\end{equation}
In practice, since $D_a$ and $SR_a$ are obtained from data 
the sum rule will not be exactly zero, but consistency requires it to
cancel within uncertainties.

In Table~\ref{tab:sr} we show the results of this sum rule calculation 
using our UFD parameterizations.
Note that it is not very well satisfied, since the $\Delta_a$
is slightly above 1.2 deviations from zero.
 This small disagreement suggests that there 
is room for improvement at low energies
in the $S$-waves.
Both the direct and integral results are compatible with the experimental 
value obtained in DIRAC, Eq.\eqref{eq:dirac}, but 
this is not surprising given the very large experimental uncertainties. 

 \begin{table}[h] 
 \caption{Sum rules in $m_\pi$ units.
 We show results for the UFD and CFD parameterizations. 
 Note that since $a_1^{3/2}$ 
 is more than 30 times smaller
 than $b_0^{3/2}$, then $D_{3/2}\sim b_0^{3/2}$.}
 \centering 
 \begin{tabular}{c | c |c } 
 \hline
 \rule[-0.15cm]{0cm}{.55cm} & UFD &  CFD  \\ 
  \hline
  \rule[-0.15cm]{0cm}{.55cm} $D_{a}$  & $0.108^{+0.018}_{-0.010}$    
&  $0.091^{+0.006}_{-0.005}$\\
  \rule[-0.15cm]{0cm}{.55cm} $SR_a$    &  $0.093\pm0.004$    &  
$0.091\pm0.003$\\
  \rule[-0.15cm]{0cm}{.55cm} $\Delta_a$  &  $0.015^{+0.020}_{-0.012}$ 
&  $0.000^{+0.006}_{-0.005}$\\
  \hline 
 \rule[-0.15cm]{0cm}{.55cm} $D_{1/2}$       &  $0.205^{+0.039}_{-0.024}$   &  $0.187^{+0.023}_{-0.016}$ \\
 \rule[-0.15cm]{0cm}{.55cm} $SR_{1/2}$      &  $0.186^{+0.006}_{-0.006}$   &  $0.182^{+0.006}_{-0.005}$ \\
 \rule[-0.15cm]{0cm}{.55cm} $\Delta_{1/2}$  &  $0.019^{+0.038}_{-0.024}$   &  $0.005^{+0.022}_{-0.016}$ \\
 \hline
 \rule[-0.15cm]{0cm}{.55cm} $D_{3/2}$       &  $-0.051^{+0.037}_{-0.005}$  &  $-0.047^{+0.005}_{-0.005}$ \\
 \rule[-0.15cm]{0cm}{.55cm} $SR_{3/2}$      &  $-0.046^{+0.003}_{-0.011}$  &  $-0.041^{+0.002}_{-0.002}$ \\
 \rule[-0.15cm]{0cm}{.55cm} $\Delta_{3/2}$  &  $-0.005^{+0.048}_{-0.007}$  &  $-0.006^{+0.006}_{-0.006}$ \\
 \hline 
 \end{tabular} 
 \label{tab:sr} 
 \end{table} 

Let us remark that a sum rule involving only scalar scattering lengths
cannot be derived 
from the $T^+$ FDR because it has one subtraction.
However, from once-subtracted FDRs it is possible to obtain 
sum rules involving scalar slope parameters and vector scattering lengths. 
Actually, by combining the $T^+$ FDR in Eq.\eqref{eq:FDRTsym} and the once-subtracted version of the $T^-$ FDR,
we can obtain two independent sum rules.
Once again we will write them as 
\begin{equation}
0=\Delta_{I}\equiv D_{I}-SR_{I},
\label{eq:DeltasrI}
\end{equation}
with $I=1/2, 3/2$. On the one hand
\begin{equation}
D_{I}\equiv b_0^{I}+3a_1^{I}
\end{equation}
will be  calculated ``directly'' from the parameterizations.
Note that $a_1^{3/2}$ 
is more than 30 times smaller
than $b_0^{3/2}$, so that $D_{3/2}\sim b_0^{3/2}$ is a very good approximation.
On the other hand, the 
$SR_{I}$ are calculated with the following integral expressions:
\begin{eqnarray}
SR_{1/2}&\equiv&
\frac{\sqrt{s_{th}}}{2 m_\pi m_K} \times\label{eq:sr12}\\
&\times&P\!\!\int^{\infty}_{s_{th}}\!\!\!\!ds'\left[\frac{\im T^+(s')+2\im T^-(s')}{(s'-s_{th})^2}\right. \nonumber \\
&-& \left. \frac{\im T^+(s')-2\im T^-(s')}{(s'+s_{th}-2\Sigma_{\pi K})^2}\right],\nonumber
\end{eqnarray}
and
\begin{eqnarray}
SR_{3/2}&\equiv& \frac{\sqrt{s_{th}}}{2 m_\pi m_K}  \times
\label{eq:sr32}\\
& \times&P\!\!\int^{\infty}_{s_{th}}\!\!\!\!ds'\left[\frac{\im T^+(s')-\im T^-(s')}{(s'-s_{th})^2} \nonumber \right. \\
&-&\left. \frac{\im T^+(s')+\im T^-(s')}{(s'+s_{th}-2\Sigma_{\pi K})^2}\right]\nonumber.
\end{eqnarray}

In Table~\ref{tab:sr} we show the results of these sum rules when the UFD set is used as input. 
As expected, the integral  result, $SR_I$, is more accurate than the direct evaluation, $D_I$, for both sum rules.
Although the fulfillment of these sum rules by our UFD set is fairly good,
this is mostly due to the large and very asymmetric uncertainties in $D_I$, 
not to a 
very good agreement in the central values.

In summary, the UFD set leaves room for improving the fulfillment of the 
sum rules just discussed.
Hence in Sec.\ref{sec:CFD} 
they will be considered, together with the FDRs,
as constraints for our parameterizations.

\section{Constrained Fits to Data}
\label{sec:CFD}

So far we have used the FDRs and sum rules as checks of our UFD set. 
We have seen that there is room for 
improvement and therefore in this section we will use them as constraints 
to obtain a new set of parametrizations, that we will call "Constrained Fit to Data" (CFD) set.
In particular we will minimize the following quantity:
\begin{equation}
W^2(d_{T^+}^2+d_{T^-}^2)+\!\!\!\sum_{I=\frac{1}{2},\frac{3}{2}}\!\!\!\left(\frac{\Delta_{I}}{\delta\Delta_{I}}\right)^2
\!\!+\sum_k \left (\frac{p_k^{UFD}-p_k}{\delta p_k^{UFD}}\right)^2,
\end{equation}
where $d_{T^\pm}^2$ are the average square distances between 
the FDR input and output defined in  Eq.\eqref{eq:distances},
$\Delta_I$ are the sum rules defined in Eq.\eqref{eq:DeltasrI}
and $\delta \Delta_I$ are their associated uncertainties.
Note that the $\Delta_a$ sum rule of Eq.\eqref{eq:SRa} is included 
in the $d_{T^-}^2$ term.
Finally,  to avoid
large deviations from the UFD data description,
we add a $\chi^2$-like penalty function
for each UFD parameter. Generically we have denoted these UFD parameters
by $p^{UFD}_k$ and their uncertainty by $\delta p^{UFD}_k$.
The $W^2=12$ constant stands for the  
number of degrees of freedom observed naively from the shape 
of $\re T^\pm$ which as seen in Fig.\ref{fig:FDRCFD} is roughly 12.
This approach is the same already followed for $\pi\pi$ scattering in
\cite{GarciaMartin:2011cn,Kaminski:2006yv,Kaminski:2006qe,Pelaez:2004vs}.

With this minimization procedure 
we have arrived to a Constrained Fit to Data (CFD) set, 
whose parameters can be found in Tables \ref{tab:S32pa} to
\ref{tab:Sinparam} and  \ref{tab:P32param} to \ref{tab:Fwave}. 
Most CFD parameters are consistent within one deviation 
with their UFD counterparts. Actually, we have allowed 46 parameters
to vary, of which 38 lie within 1 deviation,
and only three lie beyond 1.6 deviations. These are
the $\Phi_0$ parameter of the $D^{1/2}$-wave, which changes by 3 deviations, the Regge $g_{K/\pi}$ parameters, changing by 2.5 deviations and the $B_2$ parameter of the $S^{3/2}$-wave that changes by 1.8 deviations.

Before discussing in detail the changes from the UFD to the CFD set,
let us discuss first how well this new CFD set 
satisfies the FDRs and sum rules.

\subsubsection{FDRs and sum rules for the CFD set}

\begin{figure}
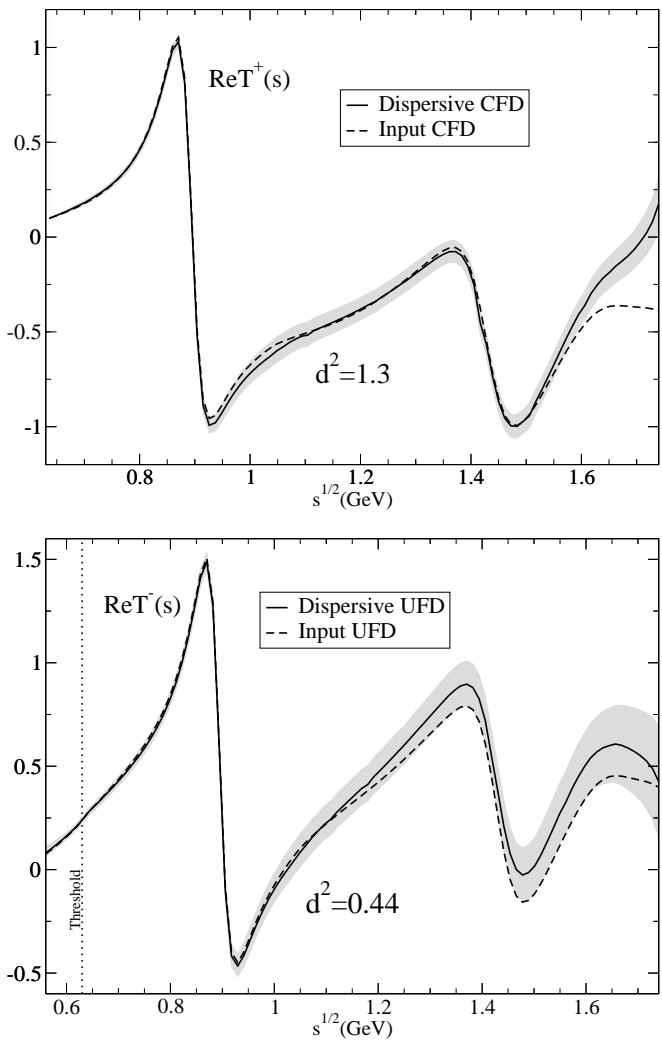

\centering
\includegraphics[scale=0.33]{Constrainedpar.eps}\vspace*{.3cm}
\includegraphics[scale=0.33]{Constrainedimpar.eps}
\caption{\rm \label{fig:FDRCFD} 
Comparison between the input (dashed) 
and the output for the $T^+$ (top) and $T^-$(bottom) FDRs when using the CFD set. 
These correspond to the left hand side versus the right hand side, respectively
in Eqs.\eqref{eq:FDRTsym} and \eqref{eq:FDRTan}.
The gray bands describe the uncertainty of the difference.
Note the dramatic improvement below 1.6 GeV compared to the
results in Fig.\ref{fig:FDRUFD} using the UFD set.
Actually, the input and dispersive CFD calculations are consistent within uncertainties
up to 1.6 GeV.
}
\end{figure}

In Fig.~\ref{fig:FDRCFD} we show the FDR results for the $T^+$ and $T^-$ amplitudes using the CFD set as input. 
These have to be compared with the corresponding results for the UFD set in Fig.\ref{fig:FDRUFD}.
Note that, in contrast to what happened when using the UFD set as input, 
 the CFD input and its dispersive output 
now agree within uncertainties. The only exception is  still the $T^+$ FDR
above 1.6 GeV, where the agreement has nevertheless improved compared to the UFD result.
For this reason in this work we only claim 
to have precise and consistent parameterizations up to 1.6 GeV.
It seems that improving the agreement above this region would require our parameterizations to depart from data. This could be due to the existence of some large systematic uncertainties in some waves or to the fact that the whole tower of higher partial waves may start to play a more relevant role.

The 
results for the averaged distances $d^2_{T^\pm}$
of the two FDRs for this CFD set are shown in Table~\ref{tab:d2FDR}. 
Let us remark that they are much
smaller than 1 below 1.6 GeV. 
The CFD set is therefore remarkably consistent
up to that energy, which is  
a dramatic improvement over the 
UFD set.
In addition, from 1.6 to 1.74 GeV the antisymmetric 
FDR is also well satisfied. 
However, although the CFD improves on the fulfillment
of the symmetric FDR above 1.6 GeV, it is still quite inconsistent.
It has not been possible to fulfill the $T^+$ FDR above 1.6 GeV
with an acceptable data description.

In Table~\ref{tab:sr} we have also provided the CFD result for the sum rules.
The central value of all $\Delta_I$ are now closer to zero
and the uncertainties are much smaller and much more symmetric.

Thus, once we have seen that the consistency of the description has improved,
let us study in detail the changes in the partial waves from the UFD to the CFD set,
and check that they still provide a good description of data up to 1.6 GeV.

\subsection{S-waves}
\subsubsection{$S^{3/2}$-wave}

\begin{figure}
\centering
\includegraphics[scale=0.33]{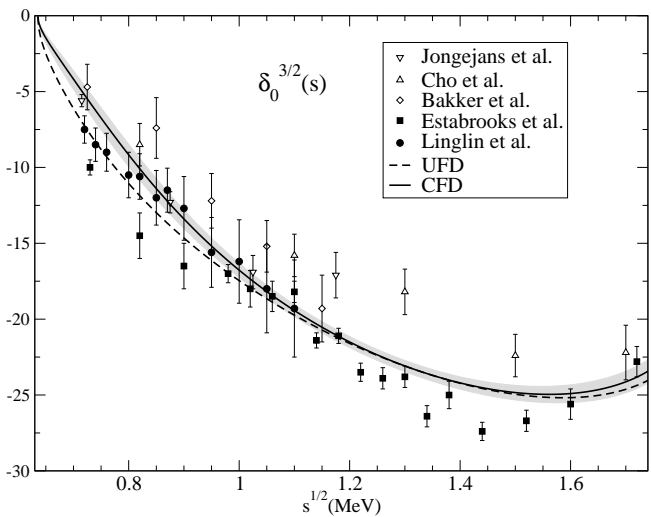}
 \caption{ \label{fig:swave32cfd} 
The CFD parameterization of the $S^{3/2}$-wave is shown as a continuous line
and its uncertainty as a gray band. For comparison the UFD parameterization is shown as a dashed line.
See Fig.\ref{fig:S32} 
for data references.
}
\end{figure}

The $S^{3/2}$-wave CFD parameters can be found in table~\ref{tab:S32pa}.
In Fig.\ref{fig:swave32cfd} we show as a continuous line the CFD 
phase shift whose uncertainties are covered by the gray band, whereas the 
UFD phase is represented by a dashed line. We do not 
plot the uncertainty band of the UFD curve 
because it was already given in Fig.\ref{fig:S32}
and  it overlaps with the CFD band.
From 1 to 1.74 GeV,
the UFD and CFD fits are almost identical, although they differ at lower energies.
In particular the central value of the
CFD scattering length is about a half of that obtained for the UFD, 
as seen in Table~\ref{tab:slengths}.

That some changes were needed at low energies in the scalar waves
was to be expected
since we already saw that the $\Delta_a$ sum rule 
was not satisfied very well by the UFD set.
Moreover, in Fig.\ref{fig:FDRUFD}  a deviation
in the low energy region of the FDRs was also observed for the UFD set.

It turns out that the FDRs and sum rule constraints 
tend to correct these small deviations by modifying only 
the $S^{3/2}$ wave at low energies.
Actually, note that both the $B_1$ and $B_2$ parameters of the $S^{3/2}$ wave
change from their UFD values by 1.5 and 1.8 deviations, respectively.
In contrast, imposing the FDR and sum rule constraints
barely changes the $S^{1/2}$-wave in the elastic region, as we will see next.
Note also that the CFD result strongly disfavors the Estabrooks et al. data at low energies. This may serve as a posteriori justification for those works that neglect these data from the start.

\begin{figure}
\centering
\includegraphics[scale=0.33]{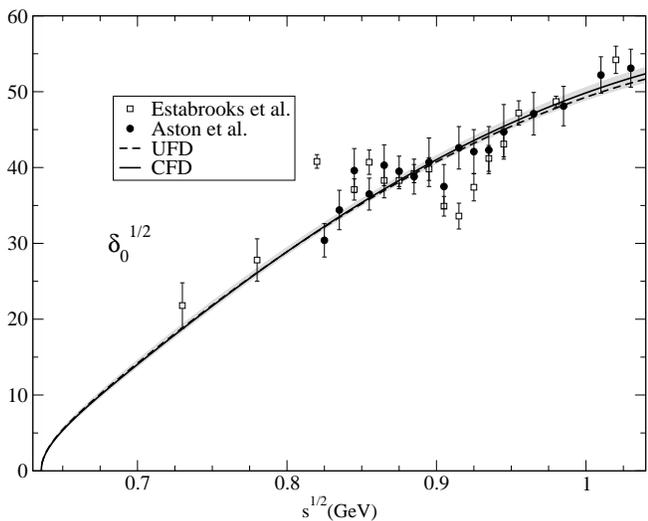}
 \caption{\label{fig:swave12cfd} 
CFD parameterization of the $S^{1/2}$-wave phase shift 
in the elastic region, shown as a continuous line whose uncertainty is covered by the gray band.
For comparison we also show as a dashed line the UFD result.
Note that in this case the UFD and CFD parameterizations are almost indistinguishable.
See Fig.\ref{Fig:Swave12} for data references.
}
\end{figure}

\subsubsection{$S^{1/2}$-wave}

As it can be seen in Fig.\ref{fig:swave12cfd}
the CFD $S^{1/2}$-wave in the elastic region (continuous line)
is almost indistinguishable from the UFD parameterization. 
Actually, as seen in Table~\ref{tab:Selparam}, the $B_0$ parameter 
does not change at all, whereas the CFD $B_1$ 
central value lies within 
less than one deviation of the UFD parameter.

 \begin{figure*}
 \centering
 \includegraphics[scale=0.66]{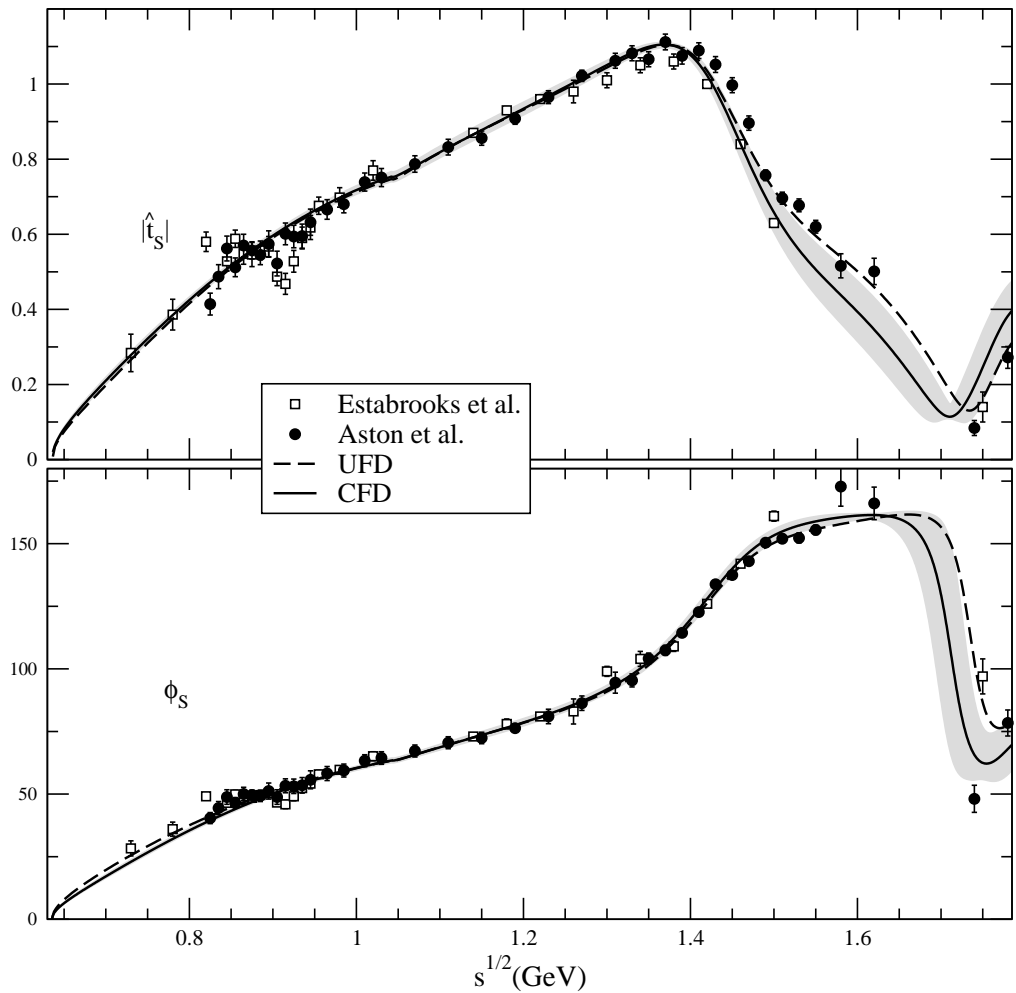}
%\vspace{-.5cm}
  \caption{\rm \label{fig:SampliCFD} 
CFD parameterization of the $\hat t_S$-wave in the whole energy region.
On the upper panel we show $\vert \hat t_S(s)\vert$ whereas in the lower one
we show $\phi_S(s)$.
The continuous line is our constrained fit (CFD), 
whose uncertainties are covered by the gray band. 
For comparison we show the UFD result as a  dashed line. Data references as in Fig.\ref{fig:Sampli}. 
}
 \end{figure*}

\subsubsection{$t_S$-wave}

As explained in previous sections, the data in the inelastic region is presented in 
terms of the modulus and phase of the $\hat t_S$ amplitude.
As seen now in Fig.\ref{fig:SampliCFD}, 
in the inelastic region the UFD and CFD descriptions are quite compatible up to 1.5 GeV.
However,
above that energy the CFD parameterization starts deviating from the UFD result.
The UFD central value (dashed line) lies slightly outside the uncertainty 
band of the CFD set, although both uncertainty bands would always overlap
and therefore the CFD still provides a fairly good description of the data.
Actually, it can be checked in Table~\ref{tab:Sinparam} that
the parameter of the CFD inelastic fit 
that varies the most with respect to its UFD value
is $\phi_1$, which changes by merely 1.1 deviations.
Around 1.7 GeV the CFD result prefers the 
solution of Aston et al. \cite{Aston:1987ir} for the phase. 
This deviation of the CFD set from the UFD one 
reflects the fact that Forward Dispersion Relations 
are not well satisfied by the UFD set in this region, 
as we already saw in Fig.\ref{fig:FDRUFD}.
As a matter of fact, the $S$-waves and the $D^{1/2}$-wave are the ones that change most to 
improve the consistency of the FDRs above 1.5 GeV.

\subsection{P-waves}
\subsubsection{$P^{3/2}$-wave}

\begin{figure}
\centering
\includegraphics[scale=0.33]{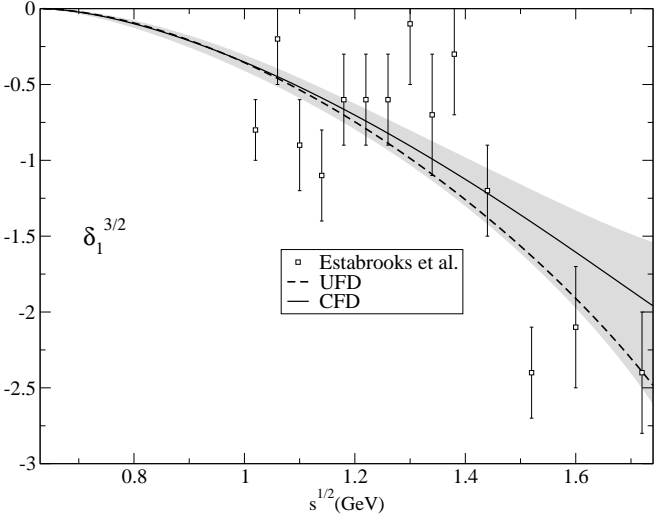}
 \caption{\rm \label{fig:pwave32cfd} 
Phase shift of the $P^{3/2}$ wave. Data from \cite{Estabrooks:1977xe}.
We show as a continuous line the CFD fit and the gray band cover its uncertainties.
The UFD result lies right on the border of this uncertainty band.
Note that this phase is rather small 
up to 1.74 GeV. }
\end{figure}

The CFD solution for this wave is shown 
as a continuous line in Fig.\ref{fig:pwave32cfd}
where its uncertainty is covered by the gray band. 
Note that the UFD solution
is compatible in the whole energy region with the new CFD parameterization. 
Moreover, in Table~\ref{tab:P32param} it can be seen that the two CFD $B_k$ parameters lie well within the uncertainties of their UFD counterparts.
Therefore the data description is still acceptable.

Let us recall that although the absolute value of this phase shift is smaller than 2.5 degrees in the whole energy region,
this wave still has some sizable effect in our calculations. This is partly due to the 
$(2l+1)$ factors
in Eq.\eqref{ec:pwexpansion}, but particularly because all other waves become 
relatively small
around 1.5 GeV.

\subsubsection{$P^{1/2}$-wave}

\begin{figure}
\centering
\includegraphics[scale=0.33]{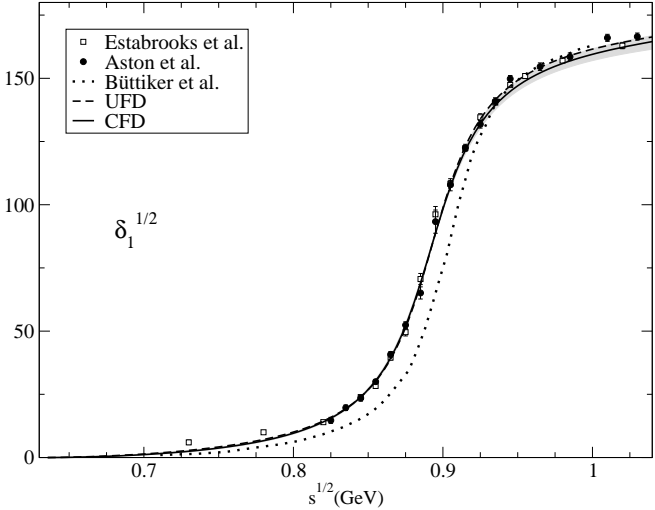}
 \caption{\rm \label{fig:pwave12cfd} 
Phase shift of the $P^{1/2}$ -wave. The CFD and UFD results are almost indistinguishable up to 
950 MeV, where the CFD phase becomes somewhat smaller. Note however that the UFD results still lies inside the uncertainty band. In addition we show the solution in \cite{Buettiker:2003pp}.
Data from \cite{Aston:1987ir,Estabrooks:1977xe}.}
\end{figure}

As seen in Fig.\ref{fig:pwave12cfd} the CFD (continuous line) 
and UFD (dashed) fits are almost indistinguishable up to 930 MeV despite the
very small uncertainty (gray band).
Around that energy, the CFD result starts deviating towards slightly lower values of
the phase, although it is still compatible with the UFD thanks to the fact that the uncertainty band
becomes larger in that region. 

This means that describing the data around the
$K^*(892)$ resonance, whose mass is $\simeq896\,$MeV and its width is $\simeq49\,$MeV, 
requires the phase in the 930 MeV to 1 GeV region to be somewhat below the existing data.
We emphasize this remark because in the 
{\it solution} of the Roy-Steiner equations in \cite{Buettiker:2003pp}, the
$K^*(892)$ phase comes out somewhat incompatible with the data (we show the result as a dotted line in Fig.\ref{fig:pwave12cfd}). To obtain such a {\it solution} the authors use as a boundary condition the value of the
phase (and its derivative) at 
$\sqrt{s}=\sqrt{0.935\,{\rm GeV^2 }}\simeq 0.967 \,{\rm GeV }$,
 which they 
take as $155.8\pm0.4^\degree$. However, at that energy, our CFD result yields $152.5\pm2.0^\degree$. This could suggest that the mismatch between the Roy-Steiner solution of \cite{Buettiker:2003pp}  and the scattering data around the $K^*(892)$ resonance could be due in part to the choice of matching phase and that
it might be improved
 by lowering it by roughly $3^\degree$, as our CFD prefers.

In the threshold region we have calculated the scattering length
directly from the CFD parameterization:
\begin{equation}
m_\pi a_1^{1/2}=0.024^{+0.008}_{-0.005}\label{eq:Psc},
\end{equation}
to be compared with the UFD result
$m_\pi a_1^{1/2}=0.031^{+0.013}_{-0.008}$.
Note that 
since our UFD and CFD fits describe the data in Fig.\ref{fig:pwave12cfd},
the resulting scattering lengths are 
larger than the one obtained in \cite{Buettiker:2003pp},
$m_\pi a_1^{1/2}=0.019\pm 0.001$.

\subsubsection{$t_P$-wave}

\begin{figure}
\centering
\includegraphics[scale=0.33]{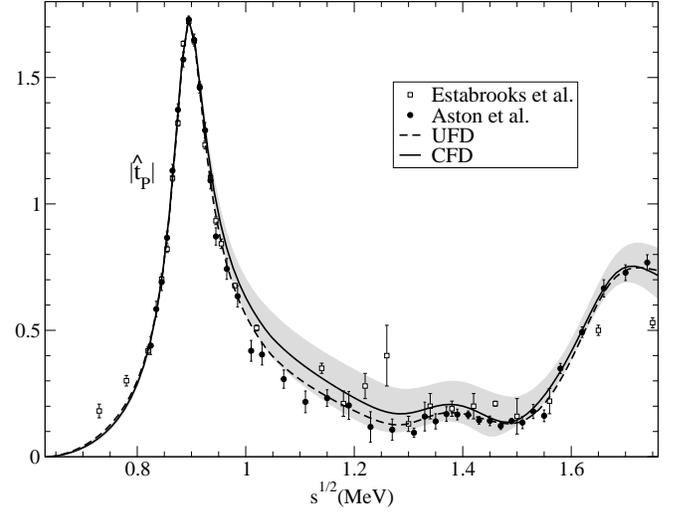}
 \caption{\rm \label{fig:pwavemodcfd} 
Modulus of the $\hat t_P=\hat t_1^{1/2}+\hat t_1^{3/2}/2$ amplitude. We show the CFD fit as a continuous line and its uncertainty as a gray band. Note that the UFD result (dashed line) is also compatible
within the CFD uncertainties. Data from \cite{Aston:1987ir,Estabrooks:1977xe}.}
\end{figure}

\begin{figure}
\centering
\includegraphics[scale=0.33]{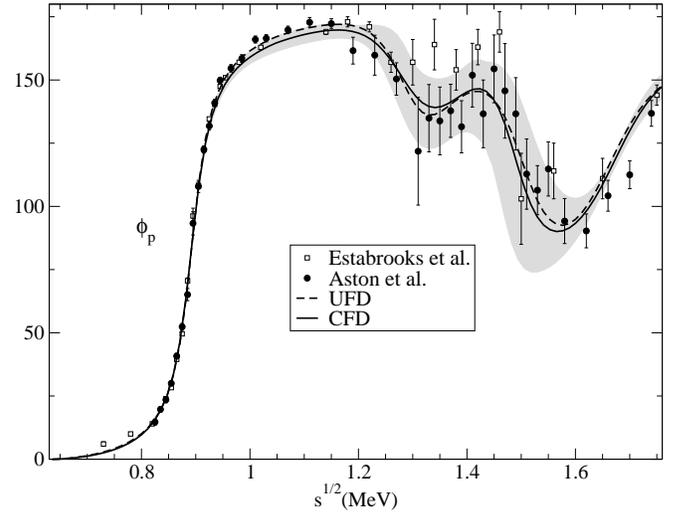}
 \caption{\rm \label{fig:pwavephasecfd} 
Phase of the $t_P=t_1^{1/2}+t_1^{3/2}/2$ amplitude. 
We show the CFD fit as a continuous line and its uncertainty as a gray band. Note that the 
UFD result (dashed line) is also very compatible
within the CFD uncertainties. Data from \cite{Aston:1987ir,Estabrooks:1977xe}.}
\end{figure}

Once we have described both the isospin $1/2$ and $3/2$ $P$-waves,
we show the modulus and phase of the 
$t_P=t_1^{1/2}+t_1^{3/2}/2$ amplitude in Fig.~\ref{fig:pwavemodcfd} and Fig.\ref{fig:pwavephasecfd},
respectively.
In the inelastic region both 
the phase and the modulus obtained for the CFD 
solution are compatible with the UFD parameterizations. 
Actually, by looking at Table~\ref{tab:Pinpa} 
one can check that the CFD parameters are almost identical to their
UFD counterparts, varying by less than one deviation, except 
the $a$ parameter, which changes by 1.4 deviations.

Our CFD solution describes fairly well the three resonances observed in this partial wave, namely the $K^*(892)$ the $K^*(1410)$ and the $K^*(1680)$.

Let us remark that although the two parameterizations are compatible, 
the CFD result prefers, for the modulus, the data of Estabrooks 
et al. \cite{Estabrooks:1977xe} between 1 and 1.5 GeV.

\subsection{D-waves}
\subsubsection{$D^{3/2}$-wave}

\begin{figure}
\centering
\includegraphics[scale=0.33]{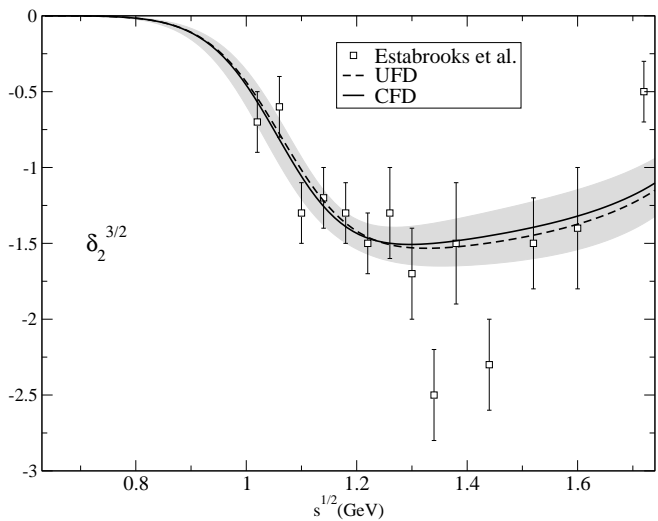}
 \caption{\rm \label{fig:dwave32cfd} 
$D^{3/2}$-phase shift. Data from \cite{Estabrooks:1977xe}.
We show the CFD result as a continuous line and its uncertainty as a gray band. 
Note it deviates very slightly from the UFD result (dashed line) and only above 1.2 GeV.
}
\end{figure}

In Fig.\ref{fig:dwave32cfd} we show the CFD result for the $D^{3/2}$-wave, whose
structure is relatively simple and its size and influence are rather small, but not completely negligible, particularly in the inelastic regime. 
As seen in the figure, the CFD solution we obtain is almost the same as the UFD parameterization. In Table~\ref{tab:D32param} it can be observed that 
the CFD parameters change by less than one third of a deviation from their
UFD counterparts.

\subsubsection{$t_D$-wave}
\begin{figure}
\centering
\includegraphics[scale=0.33]{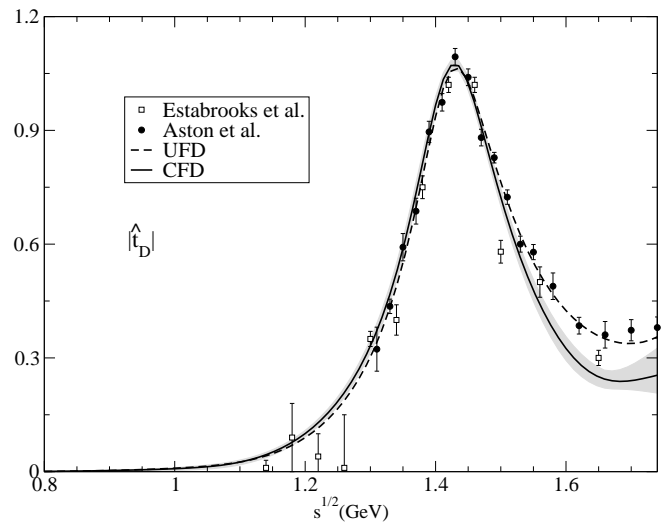}
 \caption{\rm \label{fig:dwavemodcfd} 
Modulus of the $\hat t_D$-wave. Data from \cite{Aston:1987ir,Estabrooks:1977xe}.
Note the clear peak of the $K_2^*(1430)$ resonance and that above 1.6 GeV the UFD central result (dashed line) 
is incompatible with the CFD result (continuous line) within its uncertainties (gray band).
}
\end{figure}

\begin{figure}
\centering
\includegraphics[scale=0.33]{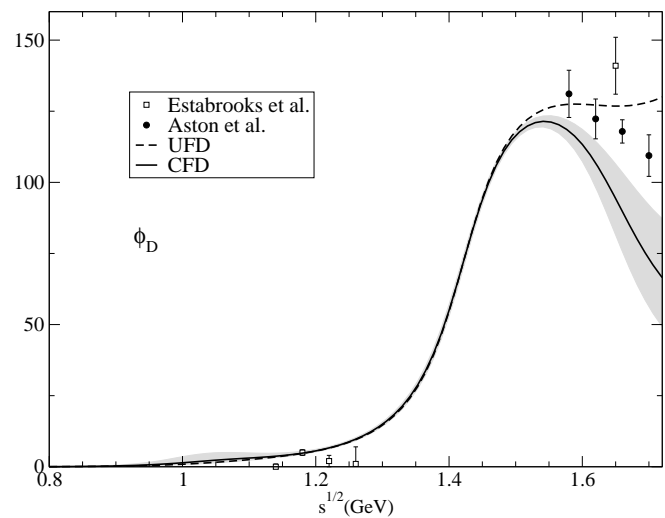}
 \caption{\rm \label{fig:dwavephasecfd} 
Phase of the $t_D$-wave. Data from \cite{Aston:1987ir,Estabrooks:1977xe}.
Note the sharp phase rise due to the the $K_2^*(1430)$ resonance and that above 1.6 GeV the existing data and the UFD central result (dashed line) are both 
incompatible with the CFD result (continuous line) within its uncertainties (gray band).}
\end{figure}

Since there are no data in the elastic region for
the $I=1/2$ $D$-wave partial wave, we directly show the modulus and phase
of the $t_D=t_2^{1/2}+t_2^{3/2}/2$ combination
in Figs.~\ref{fig:dwavemodcfd} and \ref{fig:dwavephasecfd}, respectively.
For the modulus, the CFD solution is almost indistinguishable from the UFD curve
 up to 1.6 GeV. However, above that energy
the central UFD value lies typically two to three deviations away 
from the central CFD value. Nevertheless, both fits are 
still compatible due to the 
rather large  uncertainty band of the UFD set, shown in Fig.\ref{fig:dwavemod}.
Concerning the phase, this is the curve where, above 1.6 GeV, we find the largest 
deviation from the data and the UFD set. 
By comparing the CFD versus the UFD parameters for this wave,
given in Table~\ref{tab:Dwave},  we find that the $\phi_0$ parameter
changes by more than 3 deviations. This is the only parameter that changes 
so dramatically from its UFD to its CFD value. Note it is closely related to the background produced by the opening of the $K\eta$ channel.

This deviation is not too worrisome since it occurs at the very end of our parameterizations and outside the peak of the $K_2^*(1430)$, whose width is roughly 100 MeV. Therefore, the amplitude in that region is relatively small.
At this point it is important to recall
 that the symmetric FDR, shown in Fig.\ref{fig:FDRCFD}, is well satisfied by the CFD set
only up to 1.6 GeV. Above that energy, it improves the UFD result, but it is not enough to consider it satisfactory.
As already commented, this is one of the reasons why in this work we 
claim to have precise and consistent data parameterizations up to 1.6 GeV and not beyond. Above that region the measured data is hard to reconcile with the dispersive constraints.
This  might be due to the existence of 
further systematic uncertainties, not necessarily in this wave, 
or to the increasingly important contribution from the tower of partial waves 
to the partial wave expansion.

\subsection{$F^{1/2}$-wave}

\begin{figure}
\centering
\includegraphics[scale=0.33]{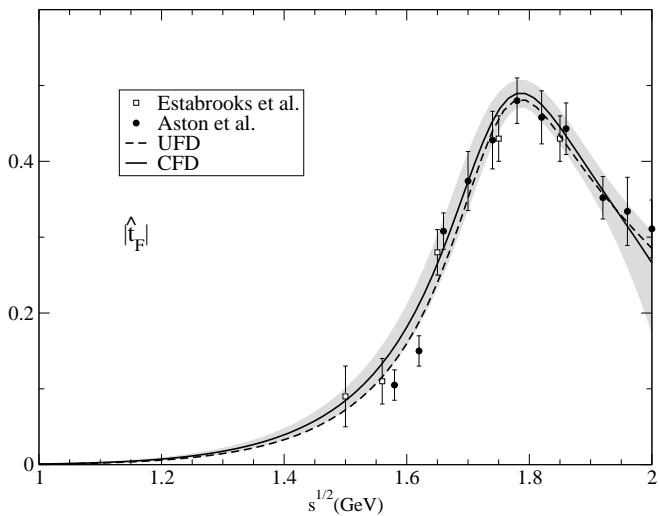}
 \caption{\rm \label{fig:fwavemodcfd}
Modulus of the $F^{1/2}$-wave. Data from \cite{Aston:1987ir,Estabrooks:1977xe}.
Note the peak of the $K_3^*(1780)$ resonance well described by
both the CFD and UFD curves, which are very compatible. The gray band stands for the CFD uncertainty.}      
\end{figure}

\begin{figure}
\centering
\includegraphics[scale=0.33]{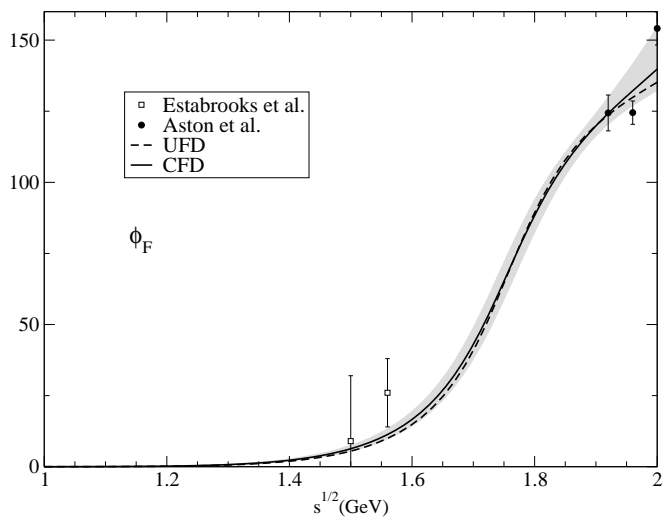}
 \caption{\rm \label{fig:fwavephasecfd} 
Phase of the $F^{1/2}$-wave. Data from \cite{Aston:1987ir,Estabrooks:1977xe}.
Note the CFD and UFD curves are very compatible. The gray band stands for the CFD uncertainty.}
\end{figure}

The CFD result for the $F^{1/2}$-wave is almost indistinguishable from
our previous UFD result and describes nicely the $K_3^*(1780)$.
This can be seen in Figs.\ref{fig:fwavemodcfd} and \ref{fig:fwavephasecfd} where we show 
the modulus and the phase of the partial wave, respectively.

\subsection{CFD Regge parameterizations}

When imposing dispersive constraints on the amplitude, we have also allowed 
the $f_{K/\pi}$, $r$ and $g_{K/\pi}$ Regge parameters to vary.
The rest of Regge parameters have been kept fixed to the values in the literature, 
also used for the UFD set and given in Table~\ref{tab:regge}. 
The reason is that,  in principle, these other parameters can be determined without using 
processes involving kaons or $\pi K$ scattering.

In Table~\ref{tab:reggeKpi} we can observe that, in the end, 
the $f_{K/\pi}$ and $r$ parameters barely change.
However the CFD value of $g_{K/\pi}$
changes by 2.5 deviations from its UFD counterpart and is responsible for more than half of the reduction
 in $d^2_{T^-}$, particularly at high energies.
As we commented before, it is not very surprising that this parameter suffers
a large change, since there is little information 
to determine it reliably. It can be considered that in this work
we are making a dispersive determination of this parameter.

\section{Discussion}

Before concluding, let us discuss our results in relation with data
obtained from the decay of heavier particles, as well as regarding poles
of resonances in the elastic regime and particularly the controversial $K_0^*(800)$ or $\kappa$-meson.

\subsection{Data from decays of heavier particles}

As already commented in Subsec.\ref{subsec:data} further information on 
the $\pi K$ system has been obtained from the decays of heavier particles.

\begin{figure}
\centering
\includegraphics[scale=0.33]{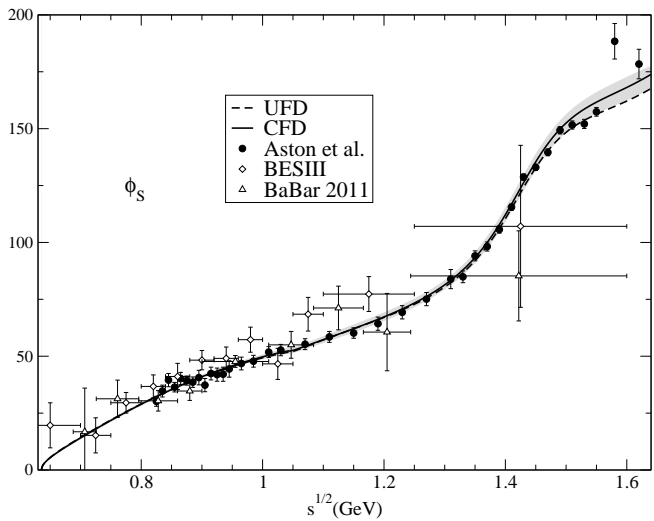}
 \caption{\rm \label{fig:semileptonic} 
Phase of the $I=1/2$ $\pi K$ $S$-wave obtained 
from the semileptonic decay $D^+\rightarrow K^-\pi^+e^+\nu_e$  
by the BaBar Collaboration \cite{delAmoSanchez:2010fd} and 
recently by the BESIII Collaboration \cite{Ablikim:2015mjo}.
These phases are compared to the LASS scattering phase shift of Aston et al.
(using their $I=3/2$ parameterization to separate the $I=1/2$).
Note that the experiments are in fairly good agreement up to 1.6 GeV.
}
\end{figure}

The semileptonic 
$D^+\rightarrow K^-\pi^+e^+\nu_e$ decays have been analyzed
by the BaBar \cite{delAmoSanchez:2010fd} and BESIII \cite{Ablikim:2015mjo} collaborations
providing 
data on the phase difference between the $S$ and $P$ components.
Since only the $\pi K$ interact strongly in the final state, 
Watson's theorem applies and in the elastic region this measurement
is nothing but the difference between the $S$ and $P$ scattering
phase shifts.
In Fig.\ref{fig:semileptonic} we show the results for the $I=1/2$ $S$-wave phase obtained
from semileptonic $D$ decays,
compared to those from scattering experiments.
Note that the uncertainties from decays are much larger than those obtained from scattering. 
Although what is actually measured in these decays is the phase-shift difference between the $P$- and
$S$-waves, the experimental collaborations provide tables for the $S$-wave
alone, by using a simple $P$-wave description, whose uncertainty is much smaller and can be neglected. A similar procedure has been followed with the LASS scattering data 
of Aston et al. \cite{Aston:1987ir}
shown in Fig.\ref{fig:semileptonic} for comparison, where the $I=3/2$ component has been separated
with the Estabrooks et al. model \cite{Estabrooks:1977xe}.
The above caveats and the very large uncertainties justify our not including 
data from decays in our fits.
All in all, there is a nice qualitative agreement between different data sources and also with our UFD and CFD results that we also show in the figure.
Moreover it is reassuring to see the good agreement between our parameterizations and the decay data in the near threshold region, where no scattering data exist.

\begin{figure}
\centering
\includegraphics[scale=0.33]{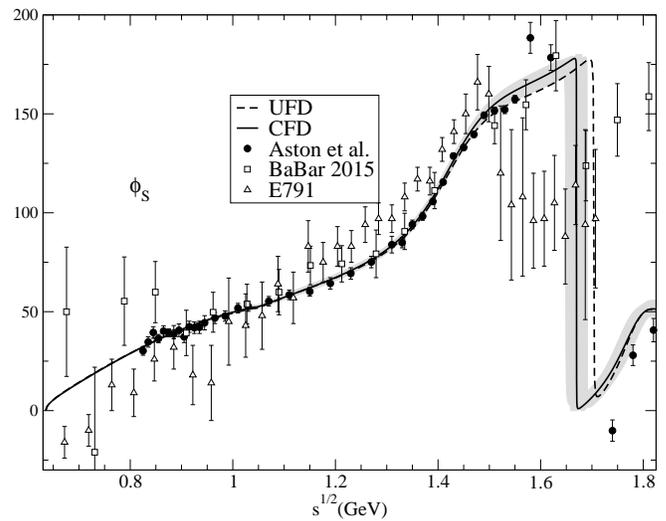}
 \caption{\rm \label{fig:fromDalitz} 
Phase of the $I=1/2$ $\pi K$ $S$-wave obtained 
from Dalitz plot analyses of $D\rightarrow K\pi\pi$  
by the E791 Collaboration \cite{Aitala:2005yh} and
of $\eta_c\rightarrow K^0_SK^\pm\pi^\mp$ by the BaBar Collaboration \cite{Lees:2015zzr}.
We have plotted the systematic plus the statistical uncertainties for \cite{Aitala:2005yh}.
These phases are compared to the LASS scattering phase shift of Aston et al.
(using their $I=3/2$ parameterization to separate the $I=1/2$).
The data from BaBar are displaced  by 34$^\degree$ 
while those from E791 are displaced by 86$^\degree$. 
Note that the qualitative agreement with the scattering phase 
only reaches up to 1.5 GeV and is not particularly good at low energies. 
}
\end{figure}

In addition, from Dalitz plot analyses it has been possible to extract the
$I=1/2$ amplitude and phase of the 
$\pi K$ $S$-wave component in $D^+\rightarrow K^-\pi^+\pi^+$ by the E791 \cite{Aitala:2005yh}, FOCUS \cite{Pennington:2007se} and CLEO-c \cite{Bonvicini:2008jw} collaborations, as well as in 
$\eta_c\rightarrow K\bar K \pi$ by the BaBar collaboration \cite{Lees:2015zzr}.
As already commented in Subsec.\ref{subsec:data}, in this case Watson's Theorem
does not imply that the phase thus measured should be the same
as that of scattering. The reason is the presence of another strongly interacting particle in the final state. This is particularly obvious by noting that the measured amplitudes 
and phases do not satisfy the elastic scattering unitarity condition.
Nevertheless, 
it has also been noticed in these works that the measured phase shows a qualitative agreement with the scattering phase-shift in the elastic region, once it is appropriately displaced by a constant. We show this qualitative agreement in Fig.\ref{fig:fromDalitz} where once again the data from scattering has been extracted using the simple $I=3/2$ model suggested by the experimental authors, which is a good enough description for a qualitative comparison. Once more, our UFD and CFD parameterizations describe well all these data.

Note, however, that the agreement disappears in the region above 1.6 GeV,
which is where we have also found that the scattering data are largely incompatible with Forward Dispersion Relations. It is then tempting to fit
in this region the phase from decays instead of the phase from scattering, in the hope that the FDRs may be better satisfied.
However, note that we can only try to fit the phase from decays, based on its similarity to the scattering phase,
but not the modulus, since the energy dependence observed for the latter is very different from that of scattering. We have performed this exercise and we have checked that the FDRs are satisfied even worse.

\subsection{Pole parameters of elastic resonances}
\label{sec:resonancepoles}

Our partial waves are constructed as piecewise 
parameterizations which are matched continuously in the real axis.
As a consequence, the resulting global amplitude does not provide a rigorous
 analytic continuation to the whole complex plane. 
Each one of the pieces may have an analytic continuation of its own, 
but at most it may only be a good approximation to the amplitude
near the part of the real axis where that particular function is used,
far from the other pieces of functions.
Nevertheless,
in the elastic region we have used a conformal mapping 
which has a well-defined analytic continuation to the complex $s$ plane.
As explained in Appendix~\ref{app:conformal}, 
the interesting feature of this mapping is that it 
places the inelastic singularities at the boundary of the unit circle.
Therefore one can expect that it 
will provide a relatively good representation of the partial wave for
complex values of $s$ which are not close to that boundary.

With these caveats in mind, we can obtain a determination of the pole positions
of resonances that appear in the elastic region 
by considering the analytic continuation 
of just the elastic conformal parameterizations. Two such resonances exist,
both with $I=1/2$, namely
the controversial $K_0^*(800)$, or $\kappa$-meson, and the $K^*(892)$ in the
scalar and vector partial waves, respectively.
Their associated poles are located in the second Riemann sheet of the partial wave,
defined as
\begin{equation}
t^{II}(s)=\frac{t(s)}{1+2i\sigma(s)t(s)},
\end{equation}
where in the upper half complex s-plane $\sigma(s)$ is defined as 
in Sec.\ref{sec:notation},
 whereas in the 
lower half plane $\sigma(s)=-\sigma(s^*)^*$.
Therefore the second sheet pole position is a solution of
\begin{equation}
\cot\delta^I_l(s_{pole})=-i,
\end{equation}
where the analytic continuation of the cotangent of the phase-shift is obtained through the conformal expansion in Eqs.\eqref{eq:cot12} and \eqref{eq:cot121} for the $K_0^*(800)$ and $K^*(892)$, respectively.

Customarily, since for narrow resonances isolated from other poles or thresholds the Breit-Wigner formula applies, one identifies the pole position of a resonance
with its mass and width as follows: $s_{R}=(M_R-i \Gamma_R/2)^2$.
Despite the $K_0^*(800)$ being a very wide resonance, we keep this convention and 
the resulting pole parameters for this resonance can be 
found in Table~\ref{tab:polesK0}, both for the UFD and CFD parameterization.
This is also the convention used in the Review of Particle Physics (RPP) \cite{RPP}.
The values we obtain are very compatible with the averaged mass in the RPP, $M_{K^*_0(800)}=682\pm29\,$MeV. In contrast, the width is somewhat larger than 
the value quoted there $\Gamma_{K^*_0(800)}=547\pm24\,$MeV. 
Actually, the most rigorous derivation is that in \cite{DescotesGenon:2006uk}
by means of a Roy-Steiner analysis, where it is found that
$M_{K^*_0(800)}=658\pm13\,$MeV and $\Gamma_{K^*_0(800)}=557\pm24\,$MeV.
Nevertheless, there is a large spread of values listed in the RPP 
and several other determinations 
find a width very similar to ours. 
As a word of caution, when comparing to the RPP
one should take into account that 
our numbers correspond to a pole position, 
whereas many values there 
correspond to peak parameterizations through Breit-Wigner 
formalisms or its variations, whose applicability is dubious due to the large width of this resonance.

\begin{table}[h] 
\caption{Pole parameters of the $K^*_0$(800) 
from the analytic continuation of the elastic parameterization only.} 
\centering 
\begin{tabular}{c c c c} 
\hline\hline  
Poles & Mass (MeV) & Width (MeV) & Coupling\\ 
\hline 
$UFD$    &  673$\pm$15  &  674$\pm$15 & 5.01$\pm$0.07 \\
$CFD$    &  680$\pm$15  &  668$\pm$15 & 4.99$\pm$0.08 \\
\hline 
\end{tabular} 
\label{tab:polesK0} 
\end{table} 

The corresponding poles for 
the vector $K^*(892)$ are found in Table~\ref{tab:polesKstar}. 
In this case the pole mass is very similar to the values provided in the RPP,
typically obtained from Breit-Wigner parameterizations. In contrast
our pole width is about 10 MeV higher that the ones listed in the RPP
or those found in $\tau^-\rightarrow K_S\pi^-\nu_\tau$ decays by the Belle collaboration \cite{Epifanov:2007rf} and on  $D^+\rightarrow K^-\pi^+ e^+\nu_e$ decays
by the BaBar collaboration \cite{delAmoSanchez:2010fd}.
It has been pointed out in \cite{Bernard:2013jxa} that this shift 
may occur on the width when fitting the LASS collaboration \cite{Aston:1987ir} phase shift
due to the fact that
those data have been given before unfolding the detector mass resolution, yielding 56 MeV instead of the 50.8 MeV quoted in the original LASS publication \cite{Aston:1987ir}.
 A similar caveat
is pointed out in Estabrooks et al. \cite{Estabrooks:1977xe}, which estimate a $\pm5\,$MeV systematic uncertainty in their width determination for this reason. 
In both cases it is pointed out that this effect barely affects the mass determination. 
Of course, all these experimental poles have been extracted
by using Breit-Wigner parameterizations modified
with Blatt-Weiskopf barrier factors, which are also model dependent.

\begin{table}[h] 
\caption{Pole parameters of the $K^*$(892) 
from the analytic continuation of the elastic parameterization only.} 
\centering 
\begin{tabular}{c c c c} 
\hline\hline  
Poles & Mass (MeV) & Width (MeV)  & Coupling\\ 
\hline 
$UFD$    &  893$\pm$1  &  56$\pm$2  & 5.95$\pm$0.07 \\
$CFD$    &  892$\pm$1 &  58$\pm$2   & 6.02$\pm$0.06 \\
\hline 
\end{tabular} 
\label{tab:polesKstar} 
\end{table} 

For the future, we plan to impose consistency 
with partial wave dispersion relations starting from the parameterizations we have obtained in this work. Those dispersion relations will provide a rigorous analytic continuation to the complex plane and a rigorous and precise determination of the resonance poles. In addition, we plan to use 
a simpler but model-independent method, recently proposed
to extract the poles from the knowledge of scattering
data in the real axis by means of Pad\'e approximants and Montessus's theorem
\cite{Masjuan:2013jha}. These two approaches are beyond the scope of the present work, which is only focused on 
obtaining a data description consistent with FDRs.

\section{Conclusions and outlook}

In this work we have presented a set of pion-kaon scattering parameterizations, which up to 1.6 GeV describe data and simultaneously satisfy a complete set of Forward Dispersion Relations as well as three sum rules for threshold parameters. Our aim has been to make the parameterizations relatively simple and easy to implement in future theoretical or experimental applications.

On a first step we have obtained a set of Unconstrained Fits to Data (UFD), in which partial waves with different angular momentum are fitted independently. Waves with different isospin are fitted together because that is how data was originally obtained. We have paid particular attention to the estimation of uncertainties, particularly to those of a systematic nature, which are not always taken into account in the literature. In addition, for the most controversial wave we have checked some statistical tests for the consistency of our uncertainty estimates.
Above 1.74 GeV, since no data on all partial waves exist, 
we have used Regge parameterizations that were obtained 
in previous works by applying factorization to other processes 
involving nucleons, pions and kaons.

However, it is shown that, even within uncertainties, this UFD set does not satisfy well Forward Dispersion Relations and also shows some tension when used inside the threshold sum rules. In particular, above the $K\eta$ threshold the dispersive results lie typically two deviations or more away from the direct calculation when using the UFD parameterizations. Throughout the elastic region the agreement is somewhat better, but still only at the level of 1.5 deviations.

Thus, on a second step,  we have imposed the Forward Dispersion relations and the sum rules as constraints on the fit parameters. Note that the parameterizations stay the same and we only change the 
values of the parameters.
Our final result is a set of Constrained Fits to Data (CFD) that satisfies Forward Dispersion Relations 
remarkably well up to 1.6 GeV while still describing the data.
In particular, the deviations
between the CFD and UFD results have been shown to be relatively small and within the uncertainties of the UFD fit. As a consequence, the CFD set still provides a good description of data. Above 1.6 GeV, we have found that the fulfillment of the dispersive constraints 
would require large modifications of the fits that  would
spoil the data description. Thus our parameterizations describe 
the data and are simultaneously consistent with dispersive constraints only up to 1.6 GeV.

Using this CFD set we have provided a precise determination of three combinations
of scattering lengths and slope parameters. In addition, 
given that the conformal map parameterization chosen for the elastic region 
has very good analytic  properties in the complex plane
we have obtained the pole 
parameters of the resonances that appear in that region, namely the vector $K^*(892)$ and the controversial scalar $K_0^*(800)$ or $\kappa$-meson. The poles and residues come in reasonably good agreement with previous determinations, although of course, the analytic continuation is dependent on our choice of conformal mapping, which is very reasonable, but not entirely model-independent.
Nevertheless, we plan to use our CFD results in the real axis as input to extract pole parameters using model-independent analytic approaches.

For the future we also plan to constraint further our parameterizations 
with a complete set of
equations of equations of the Roy-Steiner type. These are much more complicated relations written in terms of partial waves
but they are very relevant to impose crossing in addition to analyticity. Also, being formulated in terms of partial waves, they allow for a rigorous continuation to the complex plane, independent of the parameterizations used in the real axis. Thus they can provide a rigorous determination of the parameters of resonances.
Nevertheless, unlike the Forward Dispersion Relations used here, 
they are limited in practice to roughly the elastic region.
Moreover, equations of the  Roy-Steiner type 
 use as input the amplitudes in the whole energy region, 
for which it is important to use as input the CFD set obtained here.

We also expect that the simple parameterization of all the relevant partial waves
can be of use in present and future experimental and theoretical
analysis involving pions and kaons in the final state.

\section*{Acknowledgements}

{\bf Acknowledgments} JRP and AR are supported by the Spanish project FPA2011-27853-C02-02.  We are very grateful 
to B. Moussallam for kindly providing us with his parameterizations as well as for instructive comments and discussions.

\appendix

\section{Conformal expansion for elastic waves}
\label{app:conformal}

\begin{figure}
\hspace*{-.6cm}
\includegraphics[scale=0.32]{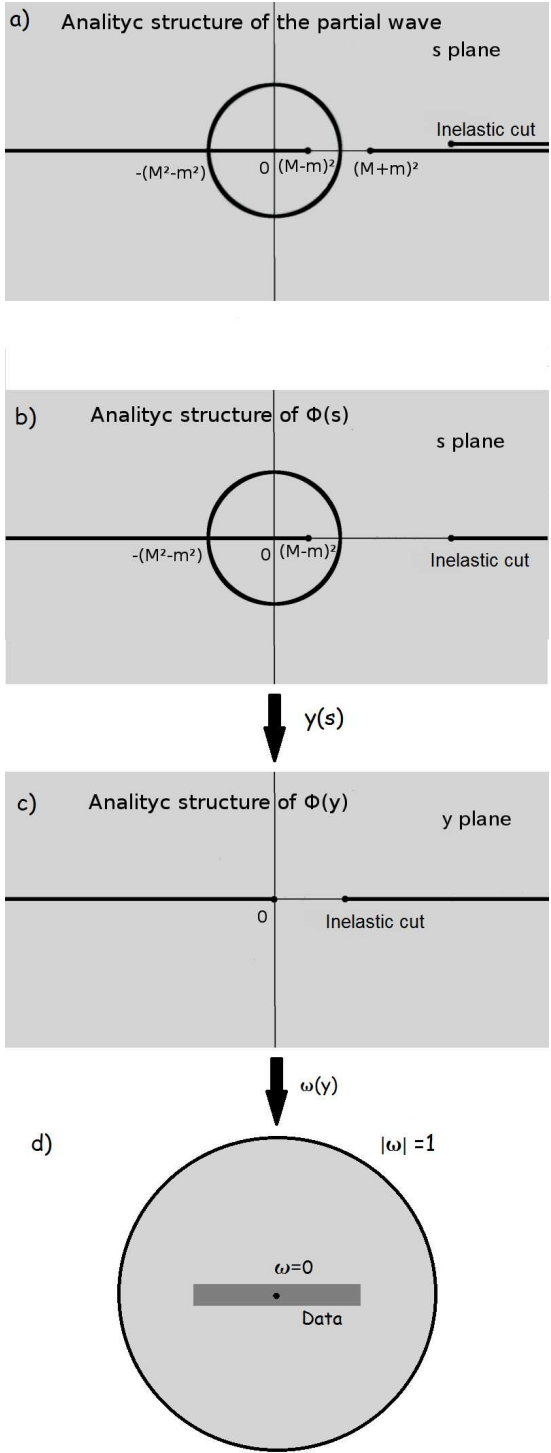}
 \caption{\rm \label{fig:cuts} 
Analytic structure in different variables
of a $\pi K$ scattering partial wave $t(s)$ and effective range function $\Phi(s)$:
a) $t(s)$ in the complex $s$-plane. Note the elastic, inelastic, left-hand and circular cuts. 
b) $\Phi(s)$ in the $s$-plane has the same structure as $t(s)$ except for the absence of the elastic cut.
c) In the $y(s)$-plane the circular cut disappears. d) The  conformal variable 
$\omega(y)$ maps the whole analyticity domain of $\Phi(y)$ inside the unit circle, 
whereas the cut singularities are confined to $\vert \omega \vert=1$.
Note that $\omega$ is defined so that the data region is roughly centered around $\omega=0$ and not too close to the border.}
\end{figure}

Let us recall that elastic partial waves can be written as
\begin{equation}
t_l(s)=\frac{1}{\sigma(s)}\frac{1}{\cot \delta_l(s)-i},
\end{equation}
where $\sigma(s)=2q/\sqrt{s}$ and $q$ is the CM momentum.
In the complex $s$-plane, partial waves for the scattering 
of two particles  with different masses $m_1$ and $m_2$ have a distinct analytic structure in the first Riemann sheet,
shown in Figure.\ref{fig:cuts}.a.
First of all, there is
a right-hand or physical cut extending from the opening of the elastic threshold to infinity. In addition, due to the thresholds in the crossed channels, there is
a left-hand cut extending from $(m_1-m_2)^2$ to $-\infty$, 
as well as  a circular cut at $\vert s \vert^2=(m_1^2-m_2^2)^2$. Other singularities may appear on the real axis when bound states exist in the direct or crossed channels, but this is not the case in $\pi K$ scattering. 
Let us emphasize that there are no poles in the first Riemann sheet.
The
cut singularities are reproduced in the second Riemann sheet, where poles can now occur anywhere in the complex plane. When poles are sufficiently close to the real axis, they give rise to resonant phenomena.

Now, in order to describe the amplitude in the complex $s$-plane,
it is customary to recast the partial wave as
\begin{equation}
t_l(s)=\frac{q^{2l}}{\Phi(s)-iq^{2l}\sigma(s)},
\end{equation}
so that, as shown in Fig.\ref{fig:cuts}.b, the effective range function $\Phi(s)$ does not have elastic cut, but only the left-hand and circular ones, as well as the inelastic cuts. 
Depending on the dynamics, it might also have  poles at the zeros of the amplitude, as we will discuss below.
In our case, it has no singularity from $\pi K$ threshold to the next inelastic threshold $s_0$. When the expansion of $\Phi(s)$ is made in terms of powers of $q$, the coefficient 
of the first term of the expansion is known as the scattering length, the second is the slope, etc... 
But the radius of convergence of this series, 
centered at $s=(m_K+m_\pi)^2$, is small, since the circular cut 
singularity lies rather close. The best way to use the largest possible domain of analyticity
is by changing variables by means of a conformal transformation. In this case, however,
it is convenient to perform first another change of variable which maps the circular cut into the left real axis:
\begin{equation}
y(s)=\left(\frac{s-\Delta_{K\pi}}{s+\Delta_{K\pi}}\right)^2,
\end{equation}
where $\Delta_{K\pi}=m_K^2-m_\pi^2$.
The resulting $\Phi(y(s))$ function now only has a right-hand ``inelastic'' cut and a left-hand cut, as shown in Fig.\ref{fig:cuts}.c, and then we can use the conformal variable
\begin{equation}
w(y)=\frac{\sqrt{y}-\alpha \sqrt{y_0-y}}{\sqrt{y}+\alpha \sqrt{y_0-y}},
\end{equation}
to map the cut $y$-plane into the unit circle in the $\omega$-plane.

With the exception of the minute $P^{3/2}$ and $D^{3/2}$-waves, in this work we have chosen $\alpha$ for each wave
so that the center of the conformal 
expansion $\omega=0$ corresponds 
to the intermediate point between the $\pi K$ 
threshold and the energy of the last data point that is fitted
with the conformal formula.
The reason for this choice is to ensure
that the region where data is to be fitted
lies well inside the $\omega$ circle, 
roughly centered around $\omega=0$, as shown in Fig.\ref{fig:cuts}.d. 
Actually, for the $S^{1/2}$ and $P^{1/2}$ waves, the data fitted with
the elastic formalism lie at
$\vert  \omega\vert<0.45$. However for the $S^{3/2}$-wave the data lie at $\vert  \omega\vert<0.6$.
The $P^{3/2}$ and $D^{3/2}$-waves are an exception, because their data starts at 1 GeV,
far form the $\pi K$ threshold. Thus we have chosen their $\alpha$ parameters so that the center of the conformal expansion corresponds to the intermediate
point where data exists. With this choice, the data fitted with this conformal expansion lies at $\vert  \omega\vert<0.6 $.

% Despite we fit with the elastic formalism the $S^{3/2}$ beyond the $\eta K$ threshold up to $\simeq 1.74\,$GeV, we have also kept $\omega(\sqrt{s_C})=0$.
% The reason is that otherwise, the threshold region would correspond to large values of $\omega$ and that would yield a too large uncertainty in that important region.
% In contrast, with our choice of $\sqrt{s_C}$,
% it is the large energy region 
% the one with larger uncertainties, which seems more adequate
% taking into account the evident incompatibilities of the data in that region,
% see Figure.\ref{fig:S32}. 
% Nevertheless we have tried centering the conformal expansion at different
% energies within the elastic region 
% and we have found that the central curves of the resulting fits
% agree below 1.2 GeV and despite deviating from each other above that energy, they 
% still fall within the uncertainty band of our fits.

Since with these changes of variable
the singularities now lie at $\vert \omega\vert =1$, the function has an analytic expansion $\Phi(s)=\sum_n{B_n w(s)^n}$  
convergent in the whole $\vert \omega\vert <1$ circle. 
In this way, and  in terms of $s$, the domain of analiticity of the conformal mapping extends to 
the whole complex plane outside the circular cut, 
with a left-hand cut and a right-hand cut above  the first inelastic threshold. Thus on the elastic region of the real axis
\begin{equation}
\cot\delta_l(s)=\frac{\sqrt{s}}{2q^{2l+1}}\Phi(s)=\frac{\sqrt{s}}{2q^{2l+1}}\sum_n{B_n w(s)^n},
\end{equation}
which are the expressions we have used for our elastic fits.

Finally, let us recall that due to chiral symmetry, scalar partial waves have a so-called
Adler zero below threshold, which is easily implemented in the partial waves 
by writing a pole factor in front of the $\Phi(s)$ expansion, as follows:
\begin{equation}
\Phi(s)=\frac{1}{s-s_{Adler}}\sum_n{B_n w(s)^n}.
\end{equation}
In addition, when there is a narrow well-established resonance and the phase crosses $\pi/2$ at $m_r$
it is also convenient to extract a factor out of the conformal expansion as:
\begin{equation}
\Phi(s)=(s-m_r^2)\sum_n{B_n w(s)^n},
\end{equation}
to accelerate the convergence of the fit.

\section{Statistical test on the $S$ wave}
\label{app:statistics}

Since the $S^{1/2}$-wave is the most controversial one
we have used some statistical tests to check the consistency of
our fits and the 
data obtained from \cite{Estabrooks:1977xe} and \cite{Aston:1987ir} 
for the $t_S\equiv t^{1/2}_0+t^{3/2}_0/2$ amplitude. 
As it has been explained in the main text, 
the problem with the data is the existence of 
large systematic uncertainties that we necessarily had to estimate.
Once we had these systematic uncertainties added to the statistical ones,
 we have performed the fits 
and obtained, by minimizing the $\chi^2$, 
the fit parameters and their uncertainties.
The $\chi^2$ is based on a Gaussianity assumption and one would like
to test that the resulting fit and the data are still consistent 
with it. 
For this reason we will
check the consistency of our fits by means of the
central moment statistical test, which in rather similar conditions was
suggested for $\pi\pi$ scattering in \cite{Perez:2015pea}.

Let $N$ be the number of data points, measured at energies $\sqrt{s_i}$, $i=1...N$. 
We then introduce a set of $N$ residuals 
$R_i=(P_i-f({\bf \alpha},s_i))/(\Delta{P_i})$. 
Here $P_i$ is the experimental value of the $i$-th measurement, $\Delta P_i$ its 
uncertainty (experimental and systematic) associated to that value,
and $f(\alpha,s_i)$ is the theoretical 
model  evaluated at $s_i$. The set of UFD parameters is called $\alpha$.

By assumption, this set of residuals must obey a standardized normal distribution. 
For this purpose we study the central moments of the residual distribution

\begin{equation}
\mu_{UFD,n}=\frac{1}{N}\sum_{i=1}^{N}{(R_i-R_{mean})^n},
\end{equation}
where $R_{mean}=\sum{R_i/N}$. 

We would like to compare these $\mu_{UFD,n}$ with the expected value of 
a set of $N$ data standardized Gaussian points. 
Thus, we generate $M$ samples of distributions of $N$ data points 
$R_{ik}$, $k=1,...M$,
that follow a normal Gaussian distribution, and calculate the central moments  $\mu_{nk}$
of
each sample. 
We then define the average central moment $\langle\mu_n\rangle=\sum_k^M \mu_{nk}/M$.
Similarly, we define the uncertainty in this distribution of residuals 
as the usual standard deviation:
 $\Delta \mu_n\equiv\sqrt{\left\langle \mu_n^2\right\rangle-\left\langle \mu_n\right\rangle^2 }$.

In order to compare the moments of our UFD result with those of the generated distributions,
we have to recall that  we have parameterized the $S$-wave into two regions with different
functional forms, and we have fitted two sets of observables, $\vert \hat t_S\vert$
and $\phi_S$. Therefore we have 4 different tests, which are
presented in Tables~\ref{tab:test1}, \ref{tab:test2}, \ref{tab:test3} and \ref{tab:test4}.

Actually, our procedure to estimate uncertainties has made use of these tests. 
At first we introduce as systematic uncertainties
 half of the distance between those points measured at the same energy 
which are incompatible. 
Then, we modify the systematic uncertainties of the few data points 
that cause deviations from the Gaussian behavior of the tests.   
With these modified systematic uncertainties the fit is performed again, the tests are
checked once more and the systematic uncertainties of points that cause deviations from the test are changed again. The procedure is iterated until the Gaussianity tests are well satisfied.

\begin{table}[h] 
\caption{Normality condition for $\Phi_S$ in the elastic region.
\label{tab:test1} } 
\centering 
\begin{tabular}{ccccccc} 
\hline\hline  
n & 1 & 2 & 3 & 4 & 5 & 6 \\ 
\hline 
$\mu_{UFD, n}$ & 0 & 0.8 & -0.3 & 1.6 & -1.1 & 4.2\\
$\mu_{random, n}$ & $0\pm 0$ & $1.0\pm0.2$ & $0\pm 0.4$ & $2.8\pm 1.6$ & $0\pm 4$ & $14\pm 17$\\ 
\hline 
\end{tabular} 
\end{table}

\begin{table}[h] 
\caption{Normality condition for $\vert \hat t_S\vert$ in the elastic region.
\label{tab:test2} } 
\centering 
\begin{tabular}{ccccccc} 
\hline\hline  
n & 1 & 2 & 3 & 4 & 5 & 6 \\ 
\hline 
$\mu_{UFD, n}$ & 0 & 1.1 & 0.1 & 2.4 & 0.8 & 6.7\\
$\mu_{random, n}$ & $0\pm 0$ & $1.0\pm0.2$ & $0\pm 0.4$ & $2.8\pm 1.6$ & $0\pm 4$ & $14\pm 17$\\ 
\hline 
\end{tabular} 
\end{table}

\begin{table}[h] 
\caption{Normality condition for the $\Phi_S$ in the inelastic region.
\label{tab:test3} } 
\centering 
\begin{tabular}{ccccccc} 
\hline\hline  
n & 1 & 2 & 3 & 4 & 5 & 6 \\ 
\hline 
$\mu_{UFD, n}$ & 0 & 1 & 0.2 & 3.3 & -2.1 & 18.1\\
$\mu_{random, n}$ & $0\pm 0$ & $1.0\pm 0.2$ & $0\pm 0.4$ & $2.8\pm 1.6$ & $0\pm 4$ & $14\pm 16$\\ 
\hline 
\end{tabular} 
\end{table}

\begin{table}[h] 
\caption{Normality condition for the $\vert \hat t_S\vert$ in the inelastic region.
\label{tab:test4} } 
\centering 
\begin{tabular}{ccccccc} 
\hline\hline  
n & 1 & 2 & 3 & 4 & 5 & 6 \\ 
\hline 
$\mu_{UFD, n}$ & 0 & 0.9 & 0.09 & 1.6 & 0.6 & 3.6\\
$\mu_{random, n}$ & $0\pm 0$ & $1.0\pm 0.2$ & $0\pm 0.4$ & $2.8\pm 1.6$ & $0\pm 4$ & $14\pm 16$\\ 
\hline 
\end{tabular}  
\end{table}

\end{document}